\documentclass[onecolumn,aps,prd,superscriptaddress,nofootinbib]{revtex4}

\usepackage{graphicx}
\usepackage{amsmath,amsfonts,amsthm}
\usepackage[hypertex]{hyperref}
\usepackage{subfigure}
\usepackage{times}
\usepackage{float}

\newcommand{\nn}{\nonumber}
\newcommand{\beq}{\begin{equation}}
\newcommand{\eeq}{\end{equation}}
\newcommand{\bea}{\begin{eqnarray}}
\newcommand{\eea}{\end{eqnarray}}
\newcommand{\tr}{\mathrm{tr}}

\newcommand{\Eq}[1]{(\ref{#1})}

\newcommand{\Dsl}{\slash{\!\!\!\!D}}
\newcommand{\Asl}{\slash{\!\!\!\!A}}

\newcommand{\one}{\ensuremath{\mathbf{1}}}

\newcommand{\bbbar}{\ensuremath{{b\bar b}}}
\newcommand{\ttbar}{\ensuremath{{t\bar t}}}
\newcommand{\mttbar}{\ensuremath{m_{t\bar t}}}
\def\d{{\rm d}}

\makeatletter
\def\simgt{\mathrel{\lower2.5pt\vbox{\lineskip=0pt\baselineskip=0pt
           \hbox{$>$}\hbox{$\sim$}}}}
\def\simlt{\mathrel{\lower2.5pt\vbox{\lineskip=0pt\baselineskip=0pt
           \hbox{$<$}\hbox{$\sim$}}}}
\makeatother

\begin{document}
\setlength{\unitlength}{1mm}

\title{\boldmath Explaining the t-tbar asymmetry with a light axigluon}

\author{Gustavo Marques Tavares and Martin Schmaltz}
\affiliation{Physics Department, Boston University, Boston, MA 02215}

\begin{abstract}

We propose an axigluon with mass between 400 and 450 GeV and flavor universal couplings to quarks to explain the Tevatron t-tbar forward-backward asymmetry. The model predicts a small negative asymmetry for t-tbar pairs with invariant mass below 450 GeV and a large positive asymmetry above 450 GeV. The asymmetry arises from interference between s-channel gluon and axigluon diagrams and requires a relatively weakly coupled axigluon ($g_{a} = g_{qcd}/3$). Axigluon-gluon interference does not contribute to the t-tbar cross section. New contributions to the cross section arise only at fourth order in the axigluon coupling and are very small for a sufficiently broad axigluon. Dijet measurements do not significantly constrain the axigluon couplings. We propose several possible UV completions of the phenomenological axigluon which explain the required small couplings and large width. Such UV completions necessarily contain new colored fermions or scalars below the axigluon mass and predict multi-jet events with large cross sections at the Tevatron and LHC.

\end{abstract}

\maketitle

\section{Introduction}

This paper proposes a light axigluon to explain the asymmetry observed in the production of \ttbar\ pairs at the Tevatron. The asymmetry has been observed in events where both tops decay leptonically~\cite{CDFdilepton} as well as in semi-leptonic events~\cite{Abazov:2007qb,Aaltonen:2008hc,D0afb,Aaltonen:2011kc}, and it significantly exceeds the Standard Model (SM) prediction \cite{neubert,Antunano:2007da, Bowen:2005ap, Kuhn:1998kw,Almeida:2008ug}.
Particularly striking is the mass dependent asymmetry
\bea\label{CDFasym}
A^\ttbar(m_\ttbar > 450\,{\rm GeV}) &=& 0.475 \pm 0.114  \nn \\ 
A^\ttbar(m_\ttbar < 450\,{\rm GeV}) &=& -0.116 \pm 0.153 \,,
\eea
measured at CDF~\cite{Aaltonen:2011kc}. It shows that most of the asymmetry
arises from \ttbar\ events with high invariant masses, while events
with low invariant masses may even have a negative asymmetry. 

\begin{figure}[b]
\mbox{\subfigure{\includegraphics[width=1.6in]{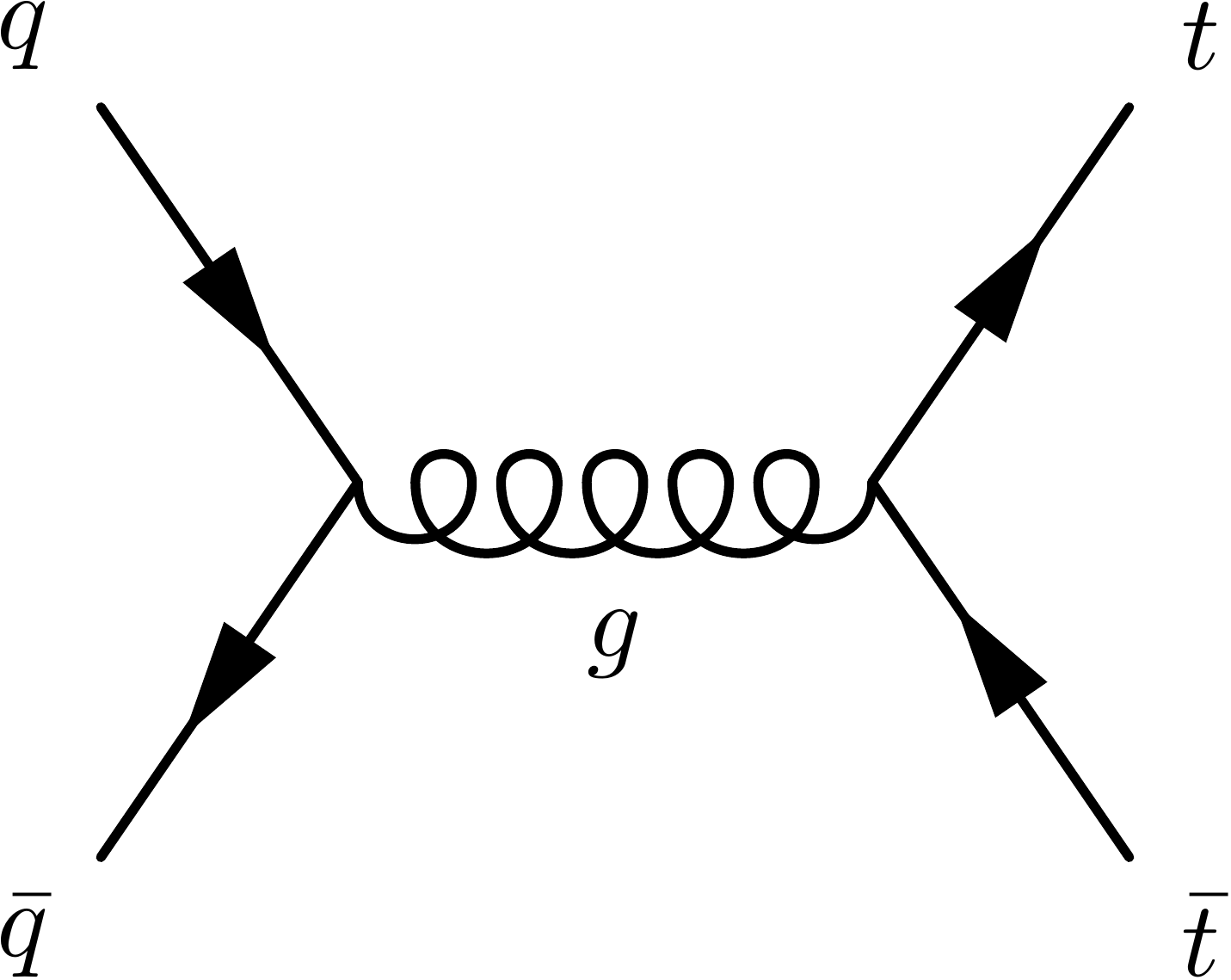}}\hskip.5in
\subfigure{\includegraphics[width=1.6in]{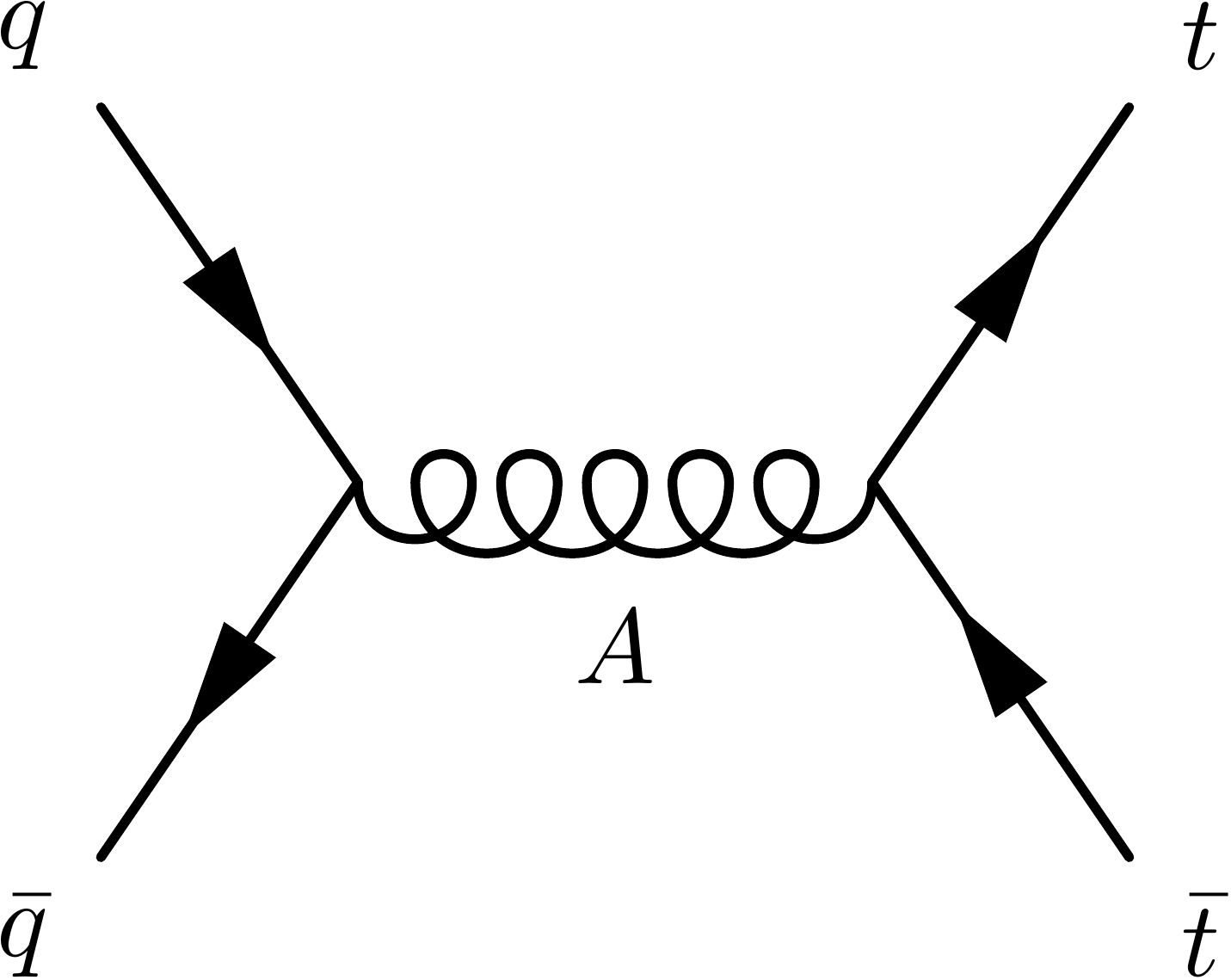}}}
\caption{S-channel gluon and axigluon diagrams.}
\label{fig:diag}
\end{figure}

A number of different models with new physics contributions to the asymmetry have been
suggested ~\cite{Frampton:2009rk,Chivukula:2010fk,Djouadi:2009nb,Djouadi:2011aj,Bauer:2010iq,Alvarez:2010js,Chen:2010hm,Bai:2011ed,Zerwekh:2011wf,Barreto:2011au,Jung:2009jz,Cheung:2009ch,Shu:2009xf,Dorsner:2009mq,Arhrib:2009hu,Barger:2010mw,Xiao:2010hm,Gupta:2010wx,Burdman:2010gr,Cheung:2011qa,Cao:2011ew,Berger:2011ua,Barger:2011ih,Grinstein:2011yv,Patel:2011eh,Jung:2009pi,Cao:2009uz,Cao:2010zb,Jung:2010yn,Choudhury:2010cd,Jung:2010ri,Blum:2011up,Craig:2011an,Delaunay:2011gv,Foot:2011xu,Ligeti:2011vt,AguilarSaavedra:2011vw,Jung:2011zv,Hewett:2011wz,Nelson:2011us,Krohn:2011tw,AguilarSaavedra:2011hz,Haisch:2011up,Cui:2011xy,Barcelo:2011vk,Gabrielli:2011jf,Duraisamy:2011pt,Shelton:2011hq,Degrande:2010kt,AguilarSaavedra:2011ug}.
Here we explore the effects of a weakly coupled axigluon~\cite{Hall:1985wz,Frampton:1987dn,Bagger:1987fz} with a mass slightly below 450 GeV. The mass is chosen to coincide with the scale $\sqrt{s}=m_{\ttbar\ }$ at which CDF observed a change-over from negative to positive asymmetry. In our model, the asymmetry arises from the axigluon-gluon interference term of the differential cross section (Figure 1). This term is proportional to the s-channel axigluon propagator
\beq 
\frac{s-M_a^2}{(s-M_a^2)^2+\Gamma_a^2 M_a^2}
\eeq
which changes sign at the mass of the axigluon $M_a$. The signs are such that the asymmetry is negative for $s<M_a^2$ and positive for $s>M_a^2$, as suggested by the CDF data. It is also interesting to consider axigluons with masses below the \ttbar\ threshold. Then the asymmetry is only very weakly $s$-dependent and positive. Motivated by the sign change of the asymmetry in the CDF data we continue to focus on values of $M_a$ between 400 and 450 GeV in this paper.
Note that our axigluon has flavor universal couplings to all quarks and therefore no constraints from flavor physics are expected.

To demonstrate that we can fit all relevant data, we compute the \ttbar\ differential cross section as a function of the axigluon mass, coupling to quarks $g_a$, and width $\Gamma_a$.
The color and spin summed and averaged squared matrix element for the process $u(p_1)\, \bar u(p_2) \to t(k_1)\, \bar t(k_2)$ is \cite{Bagger:1987fz,Frampton:2009rk}
\bea\label{matrixelem}
\frac{1}{4\cdot 9} \left|M\right|^2  =
  \frac49\, \bigg[g_s^4\, \frac{t_t^2 + u_t^2 + 2 s\, m_t^2}{s^2} 
 + 2 g_s^2\, g_a^2\, \frac{(u_t^2-t_t^2)(s-M_a^2)}{s\, ((s-M_a^2)^2+\Gamma_a^2 M_a^2)} 
  +  g_a^4\,  \frac{t_t^2 + u_t^2 - 2 s\, m_t^2}{((s-M_a^2)^2+\Gamma_a^2 M_a^2)}  \bigg] .
\eea
Here we used the partonic Mandelstam variables $s \equiv (p_1 + p_2)^2$, $t \equiv (p_1 -
k_1)^2$, $u \equiv (p_1 - k_2)^2$, and we denoted $t_t = t - m_t^2$ and $u_t = u
- m_t^2$. In terms of the top quark velocity $\beta\equiv \sqrt{1-4m_t^2/s}$ and the scattering angle $\theta$ between the outgoing top and the incoming quark in the CM frame
we have $t_t=-s(1-\beta \cos\theta)/2$ and $u_t=-s(1+\beta \cos\theta)/2$.

The second term in \Eq{matrixelem} comes from axigluon-gluon interference and is odd under the reflection $\cos\theta \leftrightarrow -\cos\theta$ ($u \leftrightarrow t$), whereas the QCD and new physics squared contributions are even. Therefore the interference term contributes to the forward-backward asymmetry but not to the differential cross section $d\sigma/dm_{\ttbar\ }$, whereas the new physics squared term contributes to the cross section but not to the asymmetry. 

The measured $p\bar p\to \ttbar$ total cross section, $\sigma_\ttbar = (7.5 \pm 0.48)$\,pb~\cite{CDFsigma} and cross section shape $\d\sigma_\ttbar/\d\mttbar$~\cite{Aaltonen:2009iz} are in reasonable agreement with predictions from perturbative QCD~\cite{Moch:2008qy,Cacciari:2008zb,Kidonakis:2008mu,Ahrens:2010zv} $\sigma_\ttbar = (6.5 \pm 0.5)$\,pb while a large new contribution to the asymmetry is required. This implies that the new physics squared term must be small for all values of $s$ while the interference term is required to be large. These two conditions are satisfied with small coupling $g_a\sim g_s/3$ and large width $\Gamma_a \gtrsim 0.1 M_a$. Much smaller values of the width would produce a noticeable ``bump'' in the \ttbar\ invariant mass spectrum while much smaller values of the coupling would fail to produce a significant asymmetry. The large values of the width which we need require additional decay channels for the axigluon beyond the decay to standard model quarks. We postpone a discussion of models which accomplish this until after showing the phenomenological fits. 

In Figure 2 we show the new physics contribution to the asymmetry as a function of invariant mass $m_{\ttbar\ }$ for three different choices of axigluon parameters. Each corresponds to an axigluon mass of $M_a=420$\,GeV and a new physics contribution to the high invariant mass asymmetry $A^{NP}(m_\ttbar > 450\,{\rm GeV}) =0.3$. Since the contributions from new physics to the differential cross section are small it is a good approximation to simply add this to the SM value of the asymmetry, $A^{SM}(m_\ttbar > 450\,{\rm GeV}) =0.11$~\cite{neubert}. Given the large uncertainties on the shape of the measured asymmetry all three are in good agreement with the asymmetry data.

\begin{figure}[t]
\includegraphics[width=4in]{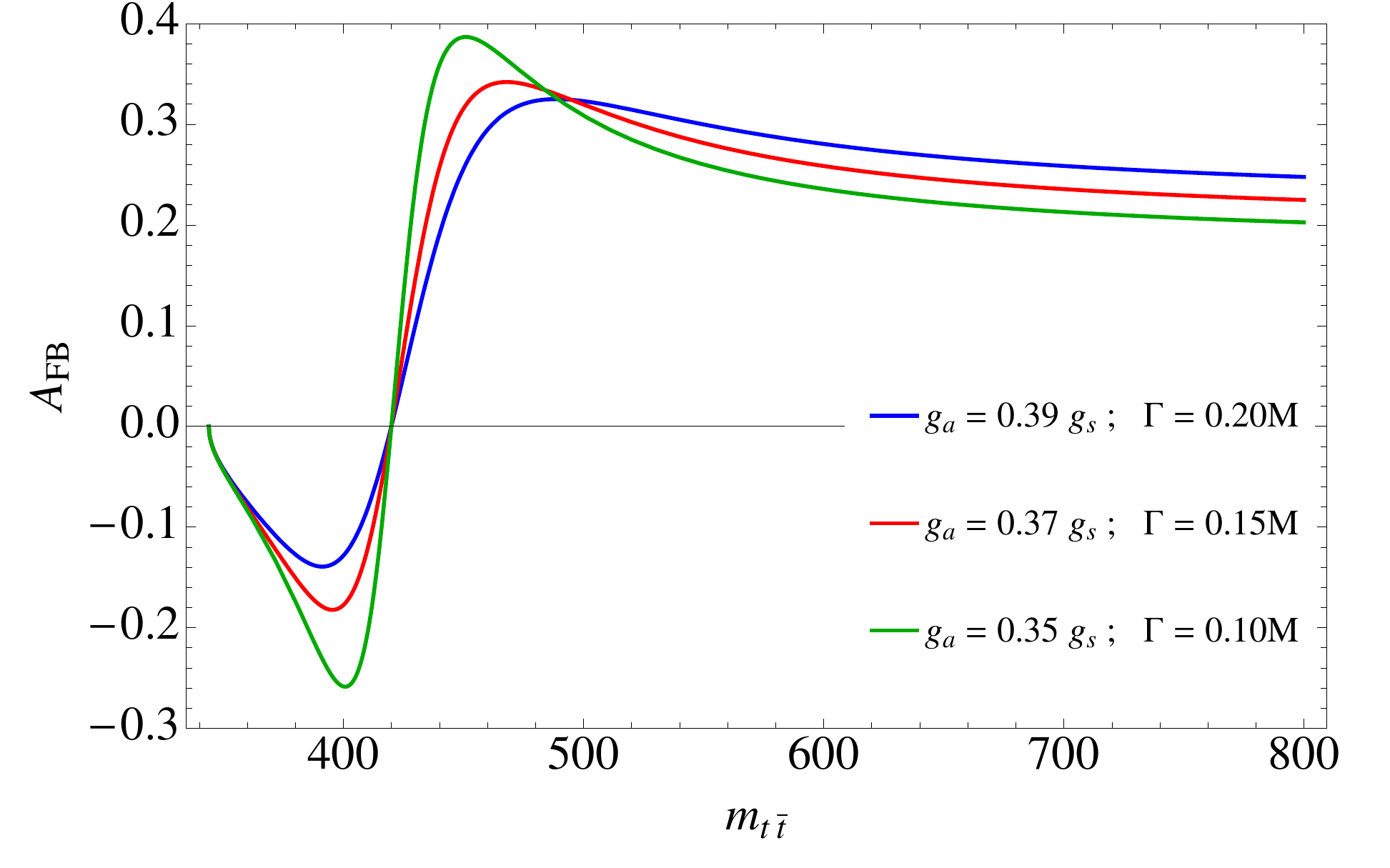}
\caption{Plot of the invariant mass distribution of the Tevatron \ttbar\ asymmetry from interference of the s-channel gluon and axigluon diagrams. The three curves correspond to axigluons with mass 420 GeV which each produce a 30\% asymmetry from new physics in the 450 GeV and above invariant mass bin. Note that the asymmetry is negative below the resonance of the axigluon. All three example points predict about -5\% asymmetry when integrated from the \ttbar\ threshold to 450 GeV. To obtain an estimate for the total new physics + QCD asymmetry, one can simply add the SM asymmetry (about 10\% in the high invariant mass bin).}
\label{fig:asym}
\end{figure}

Figure 3 shows the corresponding \ttbar\ cross sections as a function of invariant mass. One sees that for 15\% or 20\% width, the cross section shape shows very little distortion from the cross section of the SM alone. The integral of the new physics contribution under the bump in these two cases is 0.6 and 0.5 pb, respectively. This is well within the experimentally allowed cross section. 
For a width of $\simlt$ 10\% there is a visible ``bump'' in the spectrum. However, even 10\% may still be consistent with experiment after taking into account significant smearing due to detector effects and statistical fluctuations. The total new physics cross section in this case is 0.7 pb.

\begin{figure}[tb]
\includegraphics[width=4in]{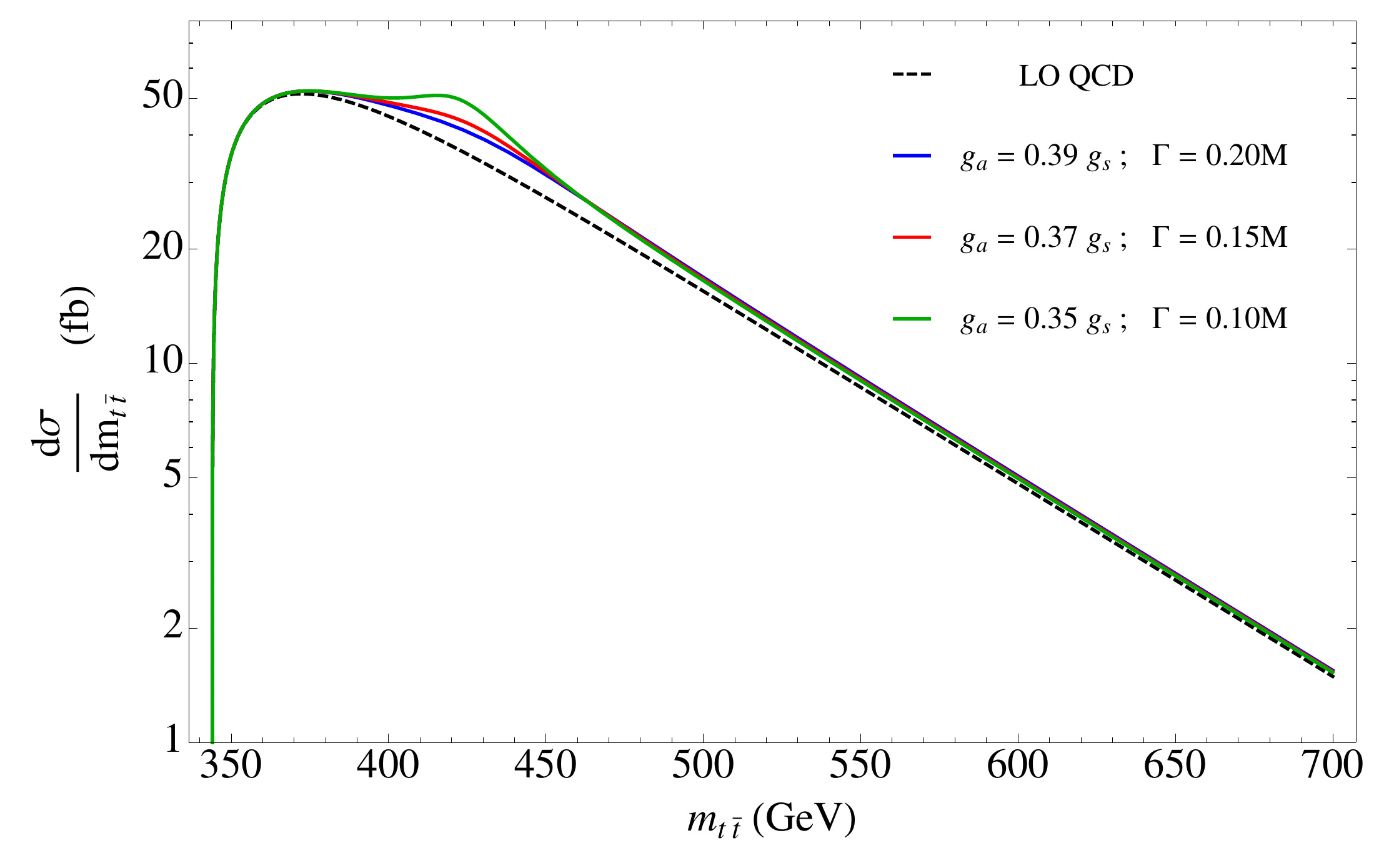}
\caption{Plot of the \ttbar\ invariant mass distribution at the Tevatron.}
\label{fig:shape}
\end{figure}

\begin{figure}[tb]
\includegraphics[width=4in]{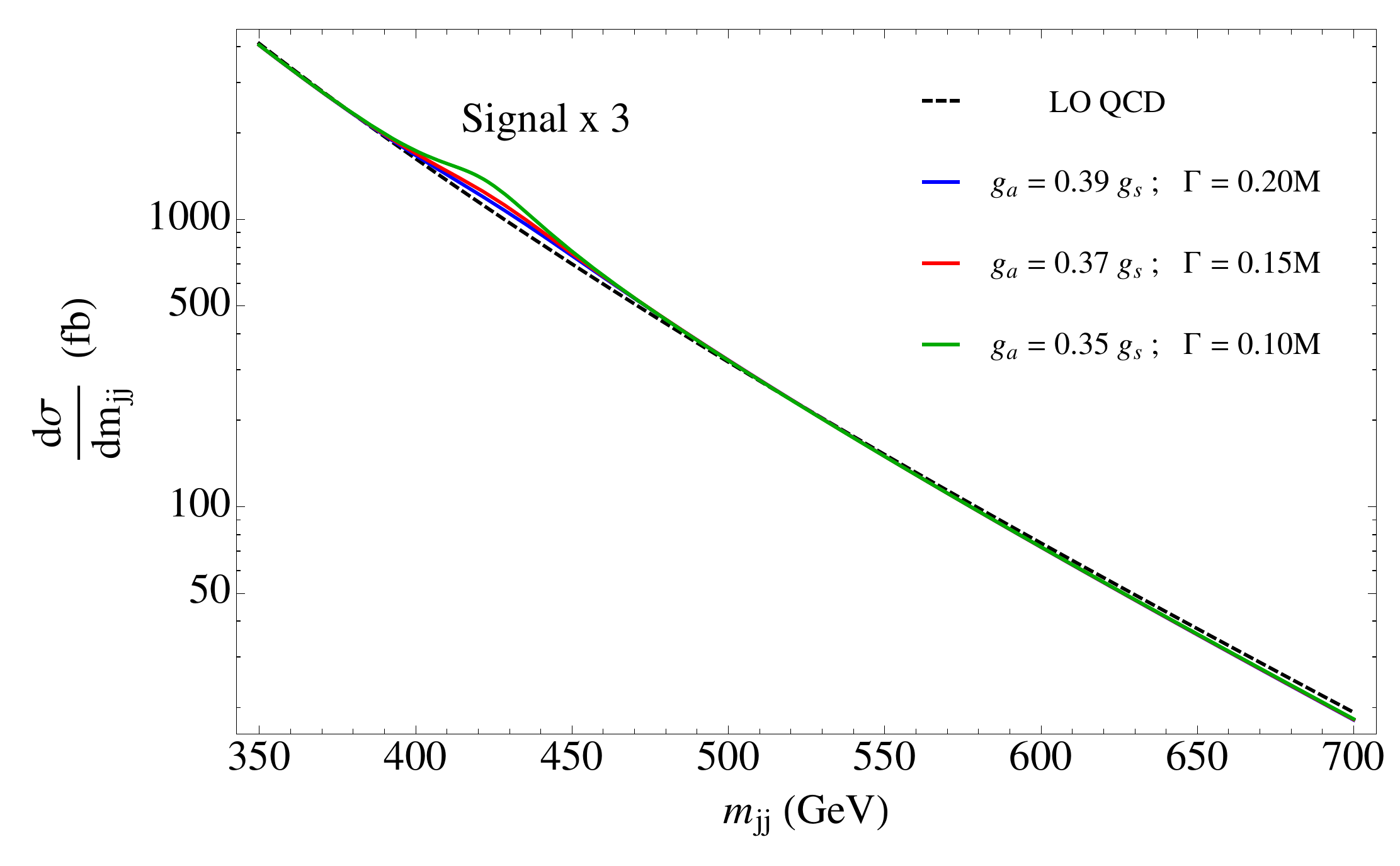}
\caption{The dijet invariant mass spectrum. We multiplied the new physics signal cross section by 3 to make it more visible.}
\label{fig:dijetshape}
\end{figure}

Another important constraint on many models comes from the absence of large deviations in the \ttbar\ cross section at the LHC~\cite{Chatrchyan:2011ew,atlasttbar} and the dijet cross sections measured at the Tevatron~\cite{Aaltonen:2008dn,CDFchidist,Abazov:2009mh} and LHC~\cite{Khachatryan:2010jd,Khachatryan:2011as,Collaboration:2010eza}. Since the axigluon in our model is relatively light and weakly coupled, the LHC top cross section does not give an interesting bound. Potentially more interesting are dijet constraints. However our axigluon is sufficiently weakly coupled and broad that the bounds are evaded provided that the new decay channels of the axigluon which are responsible for the large width do not correspond to dijets. We show a plot of the dijet invariant mass distribution including the axigluon contribution at the Tevatron in Figure~\ref{fig:dijetshape} where we multiplied the new physics contribution by a factor of 3 to make its effect visible on the plot. The integrated new physics cross section under the peak is below the CDF bound~\cite{Aaltonen:2008dn} for narrow resonances of about $8$ pb for axigluon mass $M_a=420$ GeV.

Finally, our axigluon can modify the coupling of fermions to the $W$ and $Z$ through loops. Such effects have recently been analyzed in \cite{Haisch:2011up} with the result that an axigluon as weakly coupled as ours is completely unconstrained by precision electroweak.

\section{A gauge invariant Lagrangian}

We now show how our phenomenological axigluon may be obtained from a gauge invariant Lagrangian starting with an $SU(3)_L \times SU(3)_R$ gauge group which is broken to the diagonal $SU(3)_{color}$ by the vacuum expectation value for a bi-fundamental scalar field $\phi$. The Lagrangian involves some dimension 5 and 6 couplings which we envision coming from integrating out vector-like heavy fermions with masses of several hundred GeV.  The dimension 6 couplings modify the axigluon couplings to fermions after replacing the $\phi$-field by its VEV.\footnote{A similar approach has recently been used to design low-energy couplings of an ``effective'' Z' \cite{Fox:2011qd}.}  The $SU(3)_L \times SU(3)_R$ gauge symmetry is anomalous, requiring new fermions not far above the TeV scale. We briefly discuss explicit anomaly-free UV completions in later Sections of this paper.

A good fit to the \ttbar\ cross section data requires the couplings of the axigluon to be very close to axial. This is natural if the strong interaction sector of the theory respects parity. However, parity is broken by the weak interactions and SM Yukawa coupling, and radiative corrections will generate some parity violation in the strong sector. The size of the parity violation is at least $\delta_p \sim g_2^2/16\pi^2 \log(\Lambda_{UV}/M_a)\simgt 1 \%$ which will give rise to vector couplings of the axigluon of order $\delta_p g_s$. $\Lambda_{UV}$ is model dependent and corresponds to the scale of parity breaking in the strong sector. We assume that parity violation in this sector is negligible and ignore possible small vector couplings of the axigluon. 

The Lagrangian is
\bea\label{twosu3}
{\cal L} &=& -\frac14 (F_L^a)^2 -\frac14 (F_R^a)^2 
         + Q^\dagger i \Dsl\, Q 
         + U^{\dagger} i \Dsl\, U 
         + D^{\dagger} i \Dsl\, D  \nn\\
         &+&\frac{\lambda^2}{\Lambda^2}\left[
             (\phi^\dagger Q)^\dagger\, i \Dsl\, (\phi^\dagger Q) 
         + (\phi U)^\dagger\, i \Dsl\, (\phi U) 
        + (\phi D)^\dagger\, i \Dsl\, (\phi D) \right] \nn\\
         &+& {\cal L}_{yuk} + {\cal L}(\phi)\,,
\eea
where $Q$ represents the left-handed quark doublets, $U$ and $D$ are the right-handed singlets, ${\cal L}_{yuk}$ gives rise to the SM Yukawa couplings, and ${\cal L}(\phi)$ contains the kinetic term and potential for the bifundamental scalar $\phi$. $F_L$ and $F_R$ are the field strengths for the two $SU(3)$ gauge groups, and the covariant derivatives are
\bea
D = \partial +i g\, A^{a} T^a\,.
\eea
$A$ is the $SU(3)_L$ gauge field when acting on $Q$ or $(\phi U)$ and the $SU(3)_R$ gauge field when acting on $U$ or $(\phi^\dagger Q)$, etc. The action of parity takes $F_L \leftrightarrow F_R$, $\phi \leftrightarrow \phi^\dagger$, and $Q \leftrightarrow (U,D)$.

We assume that the scalar potential forces a vacuum expectation value (VEV), $\phi = f\, \one_3$ for the scalar. The vacuum expectation value breaks two $SU(3)$ symmetries to the diagonal, and 8 Nambu-Goldstone Bosons (NGBs) are ``eaten'' by the massive axigluon. The remaining NGB from the breaking of $U(1)_L\times U(1)_R\rightarrow U(1)_V$ remains massless at this level. We can give it a small mass by breaking explicitly the off-diagonal $U(1)$ symmetry with a $\det \phi$ term in the scalar potential. Replacing the scalar with it's VEV we can solve for the mass and couplings of the axigluon by diagonalizing the gauge boson mass matrix and rescaling the fermion fields. We find the new Lagrangian
\bea\label{twosu3again}
{\cal L} &=& -\frac14 (F_V^a)^2 -\frac14 (F_A^a)^2 
         + \frac{M_a^2}{2} (A_A)^2 \nn\\
         &+& Q^\dagger \left(i \Dsl-g_a\, \Asl_A\right)\, Q 
         + U^{\dagger}  \left(i \Dsl+g_a\, \Asl_A\right)\, U 
         + D^{\dagger}  \left(i \Dsl+g_a\, \Asl_A\right)\, D  + \cdots
\eea
where now the covariant derivative contains only the gluon field and the SM weak interactions. The axigluon $A_A=(A_L-A_R)/\sqrt{2}$ couples with opposite signs to $Q$ and $U,D$. The axigluon mass, its coupling to quarks, and the strong coupling constant are
\bea
m_A=2\, g_s f\,, \hskip.3in
g_a=g_s\, \frac{1-\lambda^2 f^2/\Lambda^2}{1+\lambda^2 f^2/\Lambda^2}\,, \hskip.3in
g_s=\frac1{\sqrt{2}}\,g\,.
\eea
To obtain a small axigluon coupling $g_a$ we choose the fermion mixing parameter $\lambda f/\Lambda$ to be close to unity; thus the new fermions cannot be much heavier than the axigluon. In fact, we will be interested in the case when they are lighter.

\begin{table}[t]
\begin{tabular}{c|c|c }
%\squeezetable
Field  &  $SU(3)_L$  &  $SU(3)_R$  \\
\hline
$Q$ & $3$ & $1$ \\ 
$U, D$ & $1$ & $3$ \\
$\phi$ & $3$ & $\overline{3}$ 
\end{tabular}
\caption{Fields and representations.}
\label{tab:fields}
\end{table}

To implement the SM fermion masses in this model we must introduce Yukawa couplings of the three generations of fermions to the Higgs field. Because $Q$ and $U,D$ are charged under different $SU(3)$ gauge groups this requires insertion of the link field $\phi$. For example, the up-type Yukawa couplings could come from a coupling of the form 
\bea
\frac{\lambda_u}{\Lambda}\, Q^\dagger H \,\phi\, U \rightarrow 
\lambda_u\frac{f}{\Lambda}\, Q^\dagger H U\,.
\eea

\subsection{The axigluon width}

If the axigluon is lighter than all the other new particles in the model it can only decay to standard model fermion pairs. Then it will have a very narrow width because of the small coupling $g_a$. This is ruled out because it would produce a significant bump in the \ttbar\ invariant mass spectrum. Therefore there must be additional colored particles which are lighter than the axigluon and which have sufficiently large couplings to the axigluon.\footnote{Hiding an s-channel heavy gluon resonance in \ttbar\ production by giving it a large width was recently suggested in \cite{Barcelo:2011vk}.}

A very interesting possibility is that this role is played by the heavy vector-like fermions which we integrated out to obtain the higher dimensional operators in \Eq{twosu3}. As we will show in the next section, in a UV completion with such heavy fermions the axigluon does have large couplings to one SM fermion and one heavy fermion
\bea
g_a^{mixed} \simeq g_s\,.
\eea
Given this coupling, the width of the axigluon is
\bea
\Gamma_a=N_f \frac{(g_a^{mixed})^2}{24 \pi}M_a \left(1-\frac{M_f^2}{M_a^2}\right)^2
\left(1+\frac{M_f^2}{2M_a^2}\right) \,,
\label{eq:fermionwidth}
\eea
where $N_f$ is the number of heavy fermion partners which are lighter than the axigluon and $M_f$ is their mass. We will allow only the partners of all first and second generation quarks as well as the right-handed bottom quarks to be below the axigluon mass. Left-handed third generation and right-handed top partners must be heavier because their decay chains lead to copious production of leptons from $W$ decays. Fortunately, it is consistent with minimal flavor violation and natural to expect precisely these particles to have significantly different masses because of the large top Yukawa coupling.  Thus we will take $N_f=9$. Assuming that the 9 heavy fermions are much lighter than the axigluon so that there is no phase space suppression, one obtains a tree level axigluon width of 15\% from these decays alone. For more realistic masses of $M_f= 200$ GeV one obtains a width of 10\%. 

Of course, the fermions must then decay in a manner which is not already ruled out by existing Tevatron and LHC searches. Direct decays via  off-shell axigluons into three light quarks appear to be ruled out by recent searches for R-parity violating gluino decays~\cite{Aaltonen:2011sg,Khachatryan:multijet} up to fermion masses of 300 GeV.\footnote{Although CDF observes an intriguing and confusing  excess of events with invariant masses near the top mass.} However, if the 9th pseudo-Nambu-Goldstone-Boson $\eta_9$ from the $U(1)_L\times U(1)_R \rightarrow U(1)_V$ breaking is light, then the heavy fermions can decay into one light fermion and $\eta_9$. The coupling responsible for this decay is of order one so that this decay mode would dominate over the three body decay. Decay widths into SM fermions with mass $m_q$ and W bosons are suppressed by mixing angles $\sim m_q/M$ and are negligible except for the top quark. The pseudoscalar axion $\eta_9$ then decays into pairs of the heaviest standard model quarks for which there exists sufficient phase space. We find that $\eta_9$ masses in the range 10 GeV to 25 GeV are consistent with experiment, ensuring that it will predominantly decay to b's. The lower bound on the mass comes from upsilon decays and the upper bound ensures that the two b quarks from $\eta_9$ are reconstructed as a single boosted jet (consisting of two nearly collinear b-quarks). Thus a typical axigluon decay will result in two light jets and one ``axion jet''. We will have more to say about the phenomenology of these states in the final section. 

\subsection{Designer widths}

 Here we consider the possibility that the heavy fermion masses are above the axigluon mass. Then we must introduce additional states below the axigluon mass to produce the large width.
Since the axigluon production cross section at the Tevatron is very large (between 50 and 100 pb), we must ensure that the final states from the decays of the new particles are not already ruled out. One option for the new particles is to add $k$ color adjoint scalars $\sigma^i_{L/R}$, $i=1\cdots k$ to each of the two gauge groups. The scalars are parity mirrors $\sigma_L \leftrightarrow \sigma_R$ of each other. By choosing the multiplicity $k$ we can dial the resulting axigluon width. After the gauge symmetry breaking, we obtain the parity even/odd linear combinations $\sigma_{\pm}=(\sigma_L\pm \sigma_R)/\sqrt{2}$ with equal masses which we choose of order 100 GeV. Re-expressing the $\sigma$ gauge couplings in terms of the axigluon field and $\sigma_\pm$ we find the coupling 
\bea
g_s f^{abc} A^c_\mu (\partial^\mu \sigma_+^a \sigma_-^b
                    +\partial^\mu \sigma_-^a \sigma_+^b)
\eea
for each scalar-pseudoscalar pair. The axigluon width to decay into two scalars is
\bea \label{axiwidth}
\Gamma = m_A \, \frac{k g_s^2}{16 \pi} \left(1-\frac{4 m_\sigma^2}{m_A^2} \right)^{3/2}
\eea
Thus to get a sufficiently large axigluon with we take $k\sim 5-10$. 
To decay the scalars $\sigma_\pm$ we introduce dimension five couplings
\bea
{\cal L} = \frac{g_s^2 \eta_+ }{48 \pi^2 \Lambda} \tr \left(\sigma_+ F_{\mu\nu}F^{\mu\nu}\right)
         -  \frac{g_s^2 \eta_- }{32 \pi^2 \Lambda} \epsilon_{\mu\nu\alpha\beta} 
           \tr \left(\sigma_- F^{\mu\nu}F^{\alpha\beta}\right) \,,
\eea
which allow both $\sigma_+$ and $\sigma_-$ to decay to pairs of gluons. Such couplings are obtained from integrating out Dirac fermions with masses $M=\Lambda$ and Yukawa couplings $\eta=\eta_++i\eta_-$ to the scalars $\sigma_{L/R}$. In absence of any other significant decay channels for the $\sigma_\pm$ the axigluon would predominantly decay to four jets. Such signatures would closely resemble those recently discussed in the context of colorons arising from strong dynamics~\cite{Kilic:2008ub,Dicus:2010bm}.  We are not aware of any 4-jet searches at the Tevatron or LHC which rule out this signal. Searches for 4-jet final states with multiple b-tags~\cite{4bees,Abazov:2010ci} at the Tevatron do not apply here, and more recent searches for gluinos with R-parity violating decays resulting in six jet final states have such aggressive cuts that axigluon events would not pass selection cuts~\cite{Essig:2008zz,Aaltonen:2011sg,Khachatryan:multijet}.

\section{UV completions}

In the previous Section we presented a gauge invariant Lagrangian for the axigluon. This Lagrangian is adequate as a low-energy description of the axigluon and its interactions. However the scale suppressing higher dimensional operators cannot be very high because $g_a=g_s/3$ requires $\lambda f/\Lambda =1/\sqrt{2}$. We therefore explore a few example UV completions. 

\subsection{A minimal two site model}

\begin{figure}[ht]
\includegraphics[width=2in]{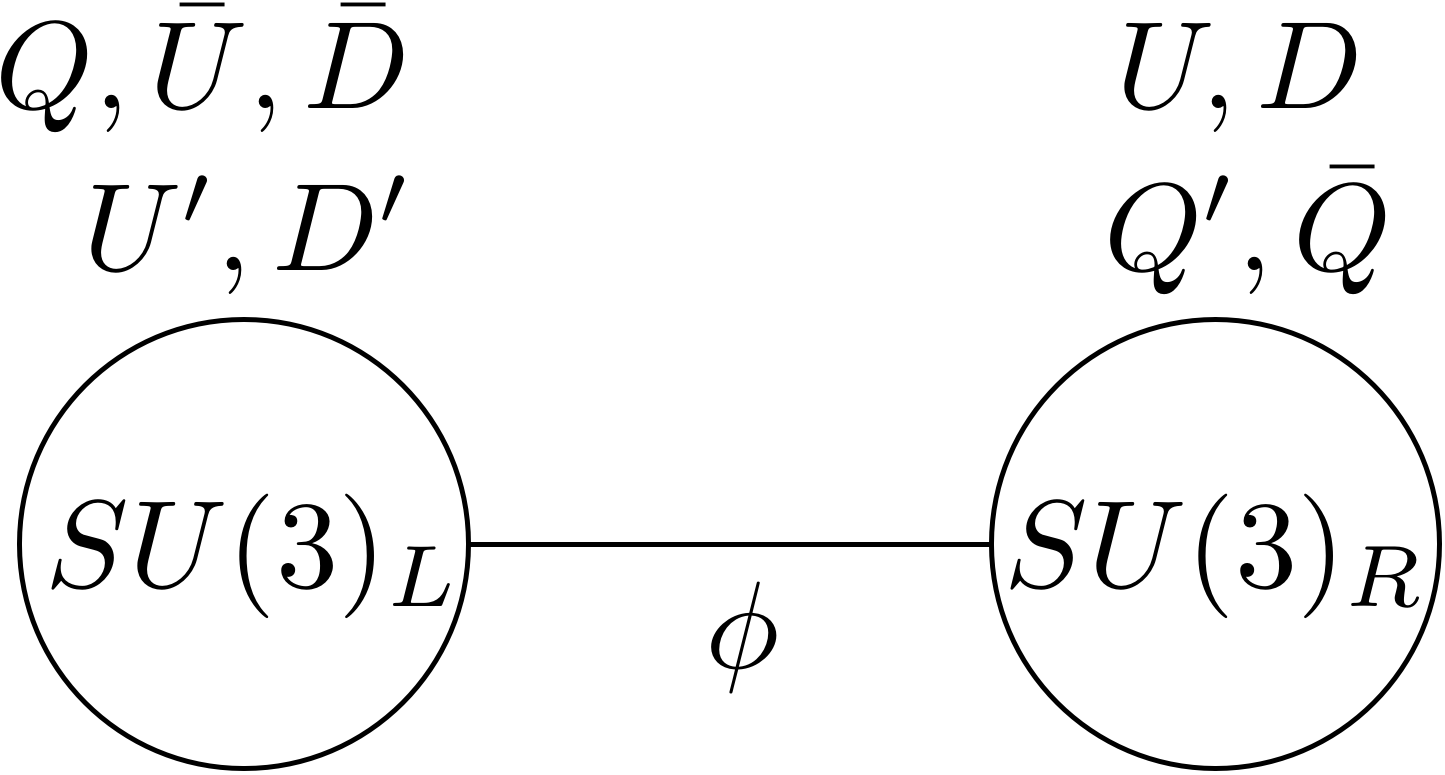}
\caption{Moose diagram for the 2 site model.}
\label{fig:twosite}
\end{figure}

The gauge group of this model is $SU(3)_L\times SU(3)_R$ spontaneously broken to $SU(3)_V$ by the VEV of a bifundamental scalar $\phi$. In addition to the fields of the previous model we introduce three generations of massive vector-like fermions $\overline{Q}+Q'$ charged under $SU(3)_R$ and $\overline{U}+U'+\overline{D}+D'$ charged under $SU(3)_L$ (see Table 2). A simple graphical representation for such a model is shown in Figure~\ref{fig:twosite}, where the gauge groups are represented by circles and the bifundamental scalar $\phi$ is a line connecting the two circles.\footnote{As described so far this model has $SU(3)_L^3$ and $SU(3)_R^3$ anomalies. These can be canceled by additional fermions $\overline Q'$ charged under $SU(3)_L$ and $Q''$ charged under $SU(3)_R$. These fermions can be given a large Dirac mass with each other after $SU(3)_L \times SU(3)_R$ breaking.}

\begin{table}[t]
\begin{tabular}{c|c|c}
Field  &  $SU(3)_L$  &  $SU(3)_R$  \\
\hline
$Q$ & $3$ & $1$ \\ 
$Q'$ & $1$ & $3$ \\ 
$\overline{Q}$ & $1$ & $\overline{3}$ \\ 
$U, D$ & $1$ & $3$ \\
$U', D'$ & $3$ & $1$ \\
$\overline{U}, \overline{D}$ & $\overline{3}$ & $1$ \\
$\phi$ & $3$ & $\overline{3}$ 
\end{tabular}
\caption{Fields and representations of the two site model.}
\label{tab:fieldsgg}
\end{table}

The fermions have the mass terms and couplings
\bea
{\cal L} &=&  \overline{Q}(M Q'+\lambda \phi^\dagger Q)
          + \overline{U}(M U'+\lambda  \phi U)
          + \overline{D}(M D'+\lambda  \phi D)\,,
\label{eq:twosite}
\eea
where flavor and parity symmetries ensure equality of the masses and couplings. 
To determine the axigluon couplings to the light fermions we may integrate out the massive fields perturbatively, expanding to second order in $\phi/M$, and treating terms with $\phi$'s as interactions. Alternatively, we may first substitute the VEV for $\phi$ and diagonalize the mass matrices for the fermions exactly. Doing the former, we would obtain the Lagrangian of the previous section~\Eq{twosu3} with $\Lambda=M$. Doing the latter, we first diagonalize the fermion mass matrices by defining 
\bea
Q_{heavy}&=&\frac{1}{\sqrt{M^2+\lambda^2 f^2}}\,(M Q'+\lambda f Q)\nn\\
Q_{SM}&=&\frac{1}{\sqrt{M^2+\lambda^2 f^2}}\,(-\lambda f Q'+ M Q)\,, 
\eea
and similar linear combinations for $U$ and $D$. The coupling of the massless linear combination $Q_{SM}$ to the axigluon is obtained by solving for $Q$ and $Q'$ in terms of $Q_{heavy}$ and $Q_{SM}$ and substituting them into the gauge kinetic terms
for $Q$ and $Q'$. We find that the axigluon couples axially to standard model and heavy quarks with coupling $g_a=g_s\, (1-\lambda^2 f^2/M^2)/(1+\lambda^2 f^2/M^2)$. There is also a coupling of the axigluon to one  standard model and one heavy quark given by
\bea
 g_s\, \frac{ 2 M \lambda f}{M^2+\lambda^2 f^2}\left(  Q^\dagger_{SM} \, \Asl_A Q_{heavy} +  Q^\dagger_{heavy} \, \Asl_A Q_{SM}  \right) - (Q\rightarrow U,D),
\label{eq:heavylight}
\eea
which, as will be discussed in the phenomenology section, can have interesting phenomenological consequences.

Note that in this model the SM Yukawa couplings can be obtained from renormalizable couplings. For example, for the up-type Yukawa couplings we may write
\bea
Y_u Q'^\dagger H U\
\rightarrow \ \  -Y_u  \frac{M \lambda f}{M^2+\lambda^2 f^2}
            Q_{SM}^\dagger H U_{SM}
\eea
As written, these Yukawa couplings break the parity symmetry and lead to small radiatively generated differences between the left and right parameters in \Eq{eq:twosite}. This leads to small vectorial couplings for the axigluon. It is possible to restore the approximate parity symmetry of the strong sector by also adding the Yukawa couplings $Y_u Q^\dagger H U'$.

The large width of the axigluon in this model derives from the decay into heavy-light fermion combinations. The coupling $g_a^{mixed}$ for this decay can be read off from \Eq{eq:heavylight}. In the limit where the axigluon coupling to the SM fermions $g_a$ becomes small this coupling approaches $g_s$, and the width is given by~\Eq{eq:fermionwidth}.

If the heavy fermions are too heavy to provide a significant width for the axigluon we must add new light particles. As in the model of the previous section we can add $k$ copies of scalars $\sigma_{L/R}^i$ with masses of order 100 GeV. The axigluon width into these particles is given by \Eq{axiwidth}. To generate the dimension 5 operators which allow the scalars $\sigma_\pm$ to decay we introduce a vector-like colored fermion for each of the gauge groups and write the couplings 
\bea
\overline \psi_L(M+(\lambda_++i\lambda_-)\sigma_L)\psi_L+
\overline \psi_R(M+(\lambda_++i\lambda_-)\sigma_R)\psi_R\,.
\eea
Integrating out the fermions generates the desired dimension 5 terms at one loop.

\subsection{The symmetric g-G-g model}

\begin{figure}[ht]
\includegraphics[width=2.5in]{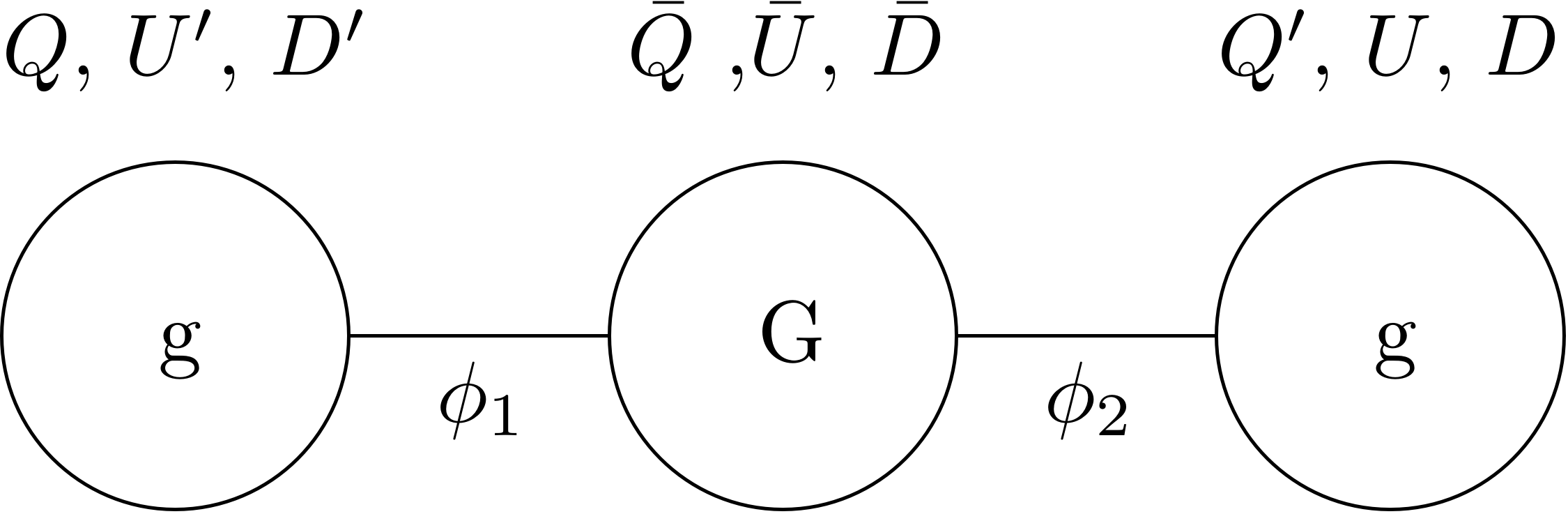}
\caption{Graphical representation for the g-G-g model.}
\label{fig:slsdiagram}
\end{figure}

This model can be described using the graphical representation of Figure~\ref{fig:slsdiagram}. There are three distinct $SU(3)$ gauge groups. The two external ones have equal gauge couplings $g$, as required by parity, and the central one has gauge coupling $G>g$. The action of parity in this model is $Q \leftrightarrow (U,D)$, $Q'  \leftrightarrow (U' , D')$, $\phi_1 \leftrightarrow \phi_2$ and $F_L \leftrightarrow F_R$.

After the link fields develop a (parity preserving) VEV $\phi_i\equiv f \, {\bf 1}_3$ there is one massless gauge boson that corresponds to the gluon, and two massive gauge bosons. One is odd under the parity transformation and is identified with the axigluon while the other is even under parity and corresponds to a ``heavy gluon''. In terms of the original parameters we find that the QCD couplings is $g_s = g G / \sqrt{g^2+2G^2}$, the axigluon mass is $M_a = g f$, and the heavy gluon mass is $M_G  = \sqrt{g^2+2G^2} \, f$. We will assume that $G \gg g$, so that the heavy gluon is much heavier and more weakly coupled to the SM fermions than the axigluon and thus does not contribute to low energy phenomenology. 

In this model the fermions that are charged under the ``external'' gauge groups have Yukawa couplings to the fermions charged under the ``central'' gauge group given by
\bea
{\cal L} = \overline Q ( \lambda_1 \phi_1^\dagger Q + \lambda_2 \phi_2^\dagger Q' ) 
+ \overline U( \lambda_1 \phi_1 U' + \lambda_2 \phi_2 U ) +(U \leftrightarrow D) + h.c.
\eea
with $\lambda_1 \sim \lambda_2$. Consequently when the scalar fields get a VEV there is a combination of $Q$ and $Q' $ that gets a mass $M_H = \sqrt{\lambda_1^2+\lambda_2^2} \, f$ with $\bar Q$ (analogously, the $U$ and $D$ fields get a mass with $\bar U$ and $\bar D$), and the other combination is identified with the standard model $Q_{SM}$. Rewriting the original fields in terms of the standard model fields and the heavy fields we find that both couple axially to the axigluon with a coupling $ g_a = g_s(\lambda_2^2-\lambda_1^2)/(\lambda_2^2 + \lambda_1^2) $. There is also a coupling of the axigluon to a light and a heavy field with strength $g_{HL} = 2 g_s \lambda_1 \lambda_2 /(\lambda_2^2 + \lambda_1^2)$, which reproduces the result from the phenomenological model for $\lambda_1/\lambda_2 \leftrightarrow \lambda f/\Lambda$. 

The SM Yukawa couplings in this model may be generated in the same way as in the model of the previous section. The large axigluon width may again be generated from decay into heavy-light fermions or from decay into additional scalars. This model is gauge anomaly free provided that $Q,U,D$ and the leptons have the usual SM $SU(2)\times U(1)$ charge assignments.

\subsection{The G-g-G model for large axigluon widths}

\begin{figure}[ht]
\includegraphics[width=2.5in]{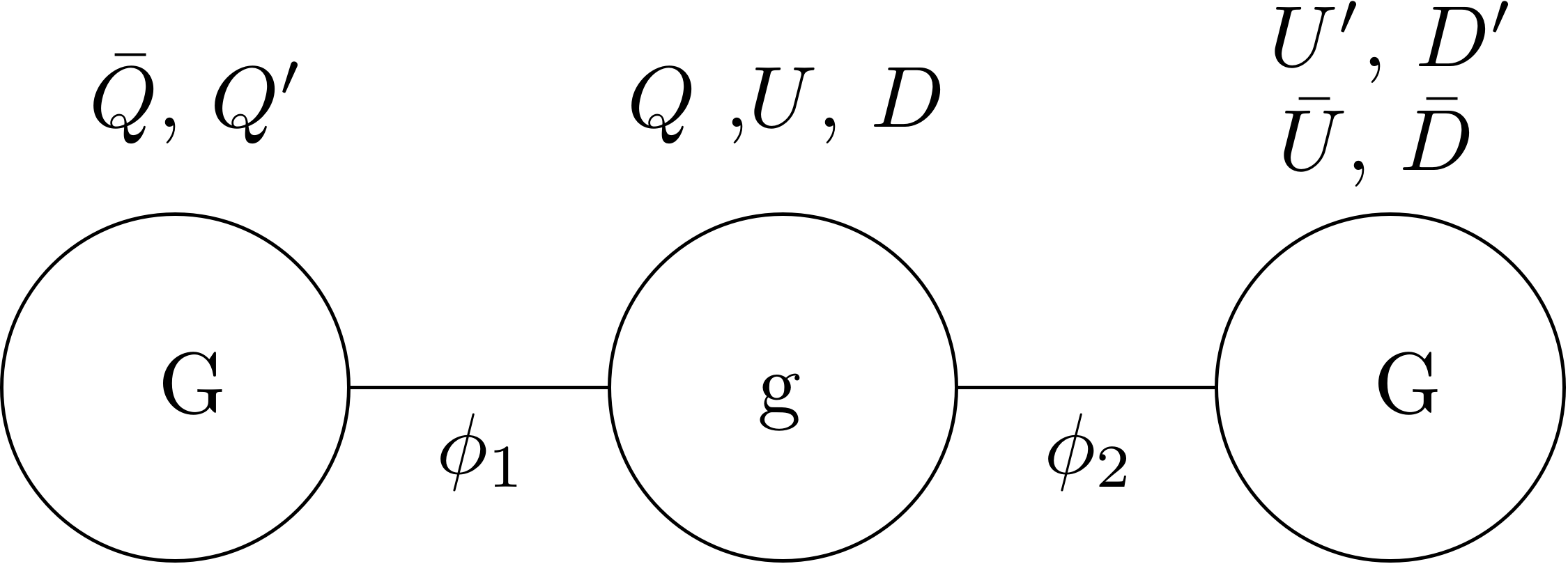}
\caption{Graphical representation for the G-g-G model.}
\label{fig:lsldiagram}
\end{figure}

This model is represented graphically in Figure~\ref{fig:lsldiagram}. It is also anomaly free. One can infer the masses of the axigluon and heavy gluon in the G-g-G model from the previous one by changing $g \leftrightarrow G$.  The QCD coupling is given by $g_s=gG/\sqrt{2g^2+G^2}$.  We include fermions charged under the central $SU(3)$ with the same quantum numbers as the SM quarks. In addition, we include heavy vector-like fermions which are charged under the external gauge groups and which mix with the fermions of the middle group. 
We will be interested in taking $G> g$. In this limit the axigluon has a large coupling to heavy fermions and can therefore have a large width. The downside is that the axigluon and the ``heavy gluon'' are approximately degenerate so that we must arrange for the heavy gluon couplings to be very small. We start with the Lagrangian
\bea
{\cal L} = \overline Q  (M Q' + \lambda \phi_1^\dagger Q) +  \overline U (M U' + \lambda \phi_2 U ) + \overline D ( M D' + \lambda \phi_2 D) + h.c.
\eea
After substituting the VEV for the scalar fields and diagonalizing the mass matrix one finds the axigluon's couplings to two SM quarks, one SM quark and one heavy quark, and to two heavy quarks
\bea
g_a^{SM}=\frac{G}{\sqrt{2}} \frac{\epsilon^2}{1+\epsilon^2}  \hspace{4pc} g_a^{mixed}= \frac{G}{\sqrt{2}} \frac{\epsilon}{1+\epsilon^2} \hspace{4pc} g_a^{heavy} = \frac{G}{\sqrt{2}} \frac{1}{1+\epsilon^2}.
\eea
Analogously, the couplings to the heavy gluon are
\bea
g_H^{SM}=\frac{G^2\epsilon^2-2g^2}{\sqrt{2G^2+4g^2}} \frac{1}{1+\epsilon^2}  \hspace{2pc} g_H^{mixed}=\frac{G^2+2g^2}{\sqrt{2G^2+4g^2}}\frac{-\epsilon}{1+\epsilon^2} \hspace{2pc} g_H^{heavy} =\frac{G^2-2g^2\epsilon^2}{\sqrt{2G^2+4g^2}} \frac{1}{1+\epsilon^2}
\eea

As desired the axigluon is weakly coupled to SM quarks if $\epsilon$ is small. The heavy gluon couplings are vectorial and have contributions from two small terms with opposite signs. In order for the model to not predict obvious features in the \ttbar\ and dijet mass spectra we must assume a cancellation of about 30\% between the two terms $G \epsilon^2-2g^2$. 

Because the quarks charged under the central gauge group have exactly the same quantum numbers as the standard model ones it is trivial to write the Yukawa couplings to the Higgs. This model is more efficient in giving the axigluon a large width because the couplings to heavy-light fermions and to pairs of heavy fermions are enhanced by factors of $1/\epsilon$ and $1/\epsilon^2$, respectively.

\section{Collider phenomenology}
\label{sec:lhc}

Before committing to a particular decay for the axigluon we can make two model independent predictions about axigluon cross sections at the Tevatron and LHC.

First, the axigluon is produced with a large cross section in the s-channel at the Tevatron. In the narrow width approximation (which is not unreasonable even at 20\% width) we expect a total axigluon production cross section of 50-100 pb at the Tevatron in the region of parameter space which can explain the \ttbar\ asymmetry. About 1\% of the axigluons contribute to a slight increase in the \ttbar\ cross section. Tevatron dijet bounds allow only about 10\% of the events to decay into dijets unless the axigluon extremely broad. Therefore most axigluons must decay into multi-jet final states for which there have not been dedicated searches. Whatever the final state, events rates so large that a dedicated search for that particular multi-jet final state would be sensitive to our signal.

Second, the axigluon as well as the colored particles which it decays into, can be pair produced with their respective QCD cross sections at the Tevatron and especially at the LHC. For example, in the interesting region of parameter space the cross section for axigluon pair production at the 7 TeV LHC is between 10 and 50 pb~\cite{Dicus:2010bm}.\footnote{Most UV completions for the axigluon model give rise to a $ g_s \tr F_{\mu\nu}[A^\mu_A, A^\nu_A] $ operator which couples two axigluons to a single gluon. The coefficient of this operator is not fixed by gauge invariance. We  assumed the tree level value for it ($\chi=1$ in the notation of \cite{Kilic:2008ub}).} Given that the axigluons decay to multi-jets we predict events with 6, 8, or even 12 jets with a cross section of 10s of pb.

\begin{figure}[t]
\mbox{\subfigure{\includegraphics[width=1.8in]{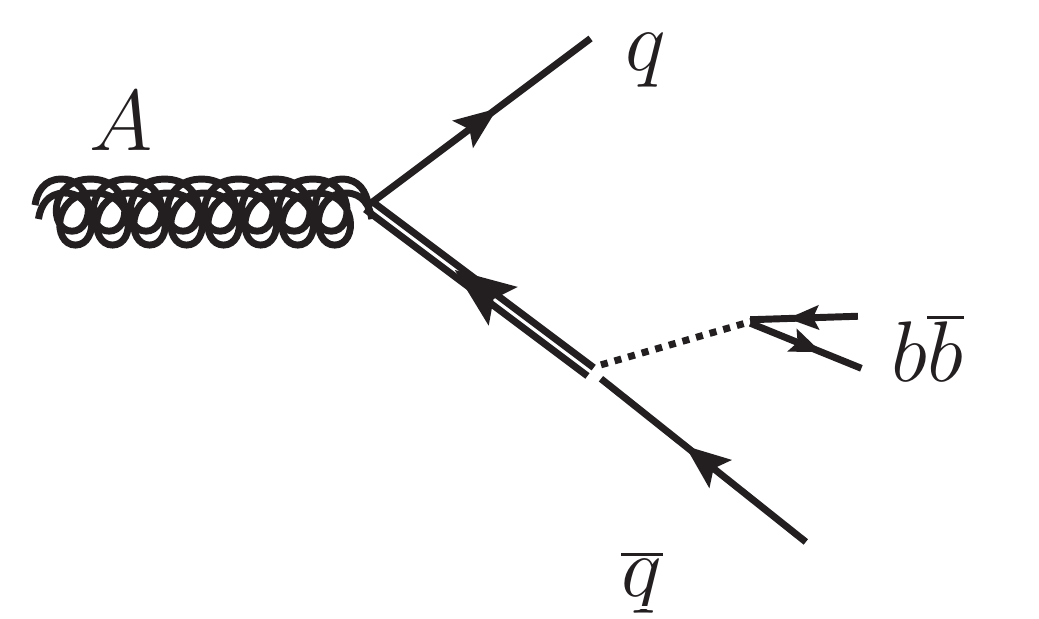}}
\subfigure{\includegraphics[width=1.8in]{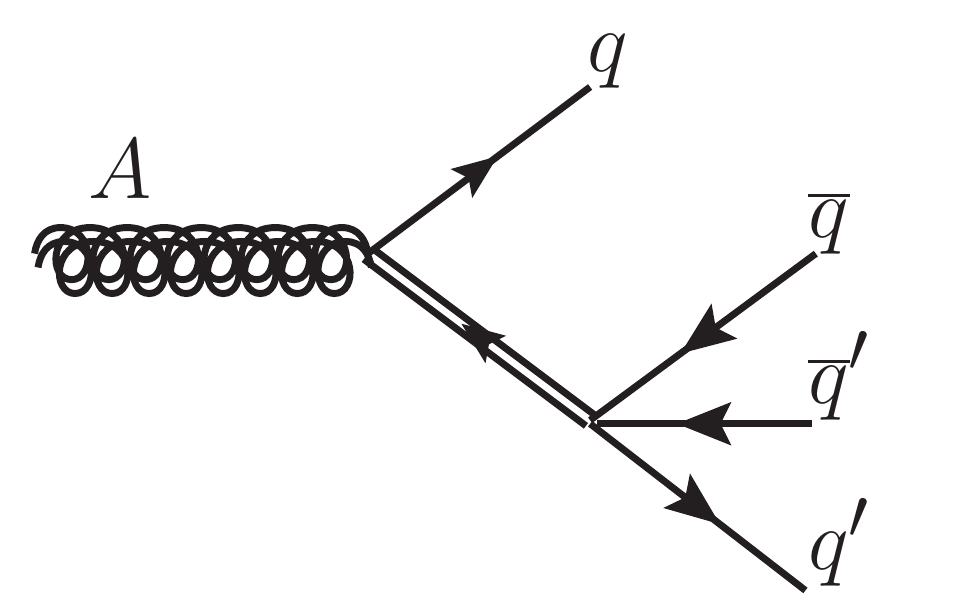}}
\subfigure{\includegraphics[width=1.7in]{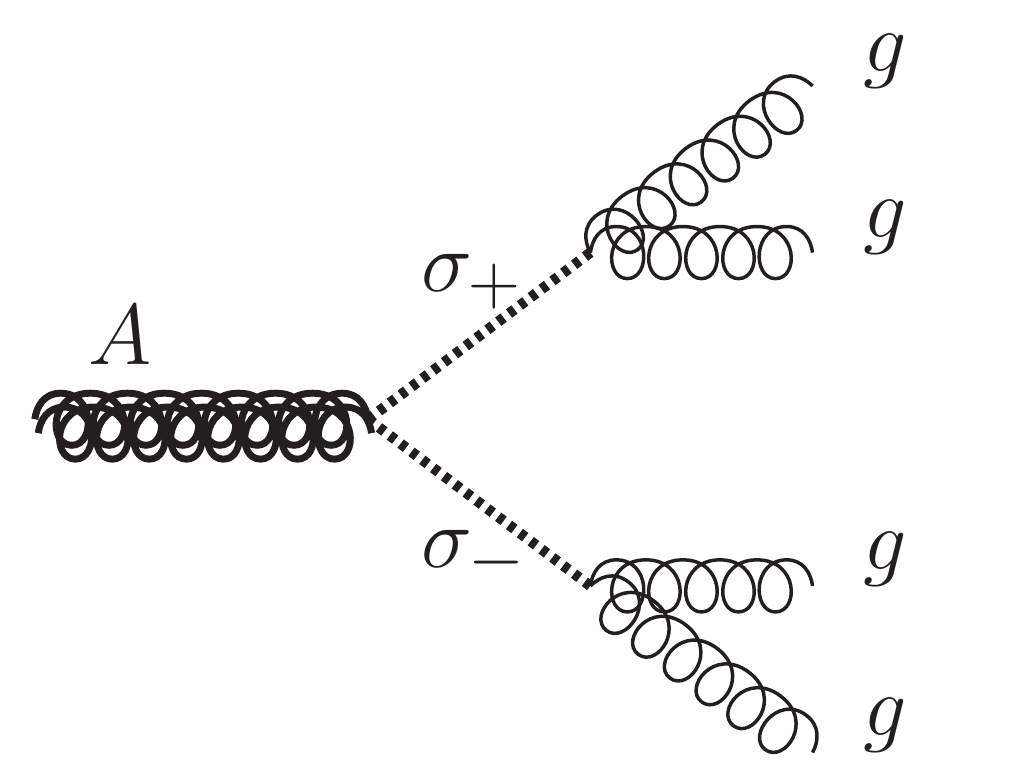}}
\subfigure{\includegraphics[width=1.6in]{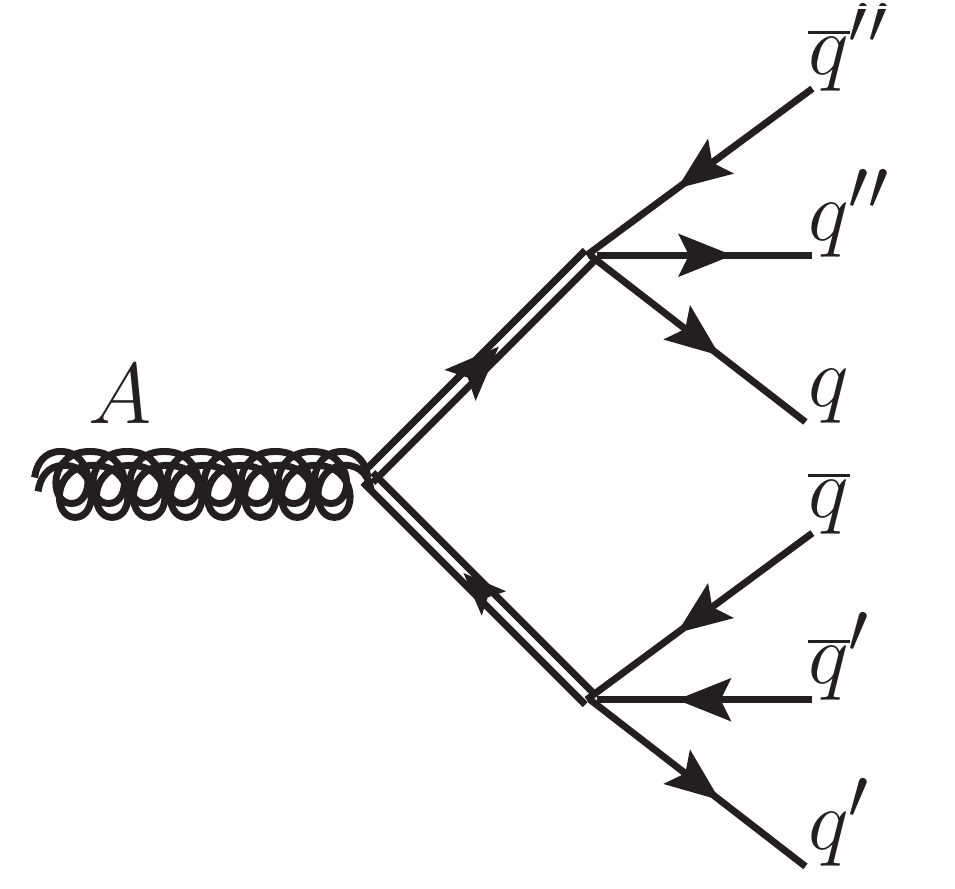}}}
\caption{Four scenarios for axigluon decay.}
\label{fig:axidecays}
\end{figure}

In the following we briefly discuss four possible scenarios for the axigluon decays. Since the production cross section at the Tevatron is so large, the axigluon would be ruled out if it had a significant branching fraction to leptons. A fifth possibility of decaying the axigluon into a pair of heavy particles which then decay into soft jets and slowly moving WIMPs appears to be already ruled out by early LHC searches~\cite{Alves:2010za}. We therefore concentrate on the four multi-jet final states depicted in Figure~\ref{fig:axidecays}.

\begin{enumerate}

\item Decay to a light quark accompanied by a heavy quark which then decays to a light quark and an axion. The axion then further decays into a boosted \bbbar\ pair  (first diagram in Fig.~\ref{fig:axidecays}). This would presumably be reconstructed as a three jet final state of which one is b-tagged. The axion jet would have a peculiar signature with very few tracks originating from the decay of a colorless particle, but it would have two displaced vertices.  One could reconstruct the total invariant mass as well as the heavy quark invariant mass at the Tevatron. At the LHC one would look for a final state with 6 jets of which two are b-tagged from axigluon pair production, or for a four jet final state with two b-tags from heavy quark pair production.

\item Decay to a light quark accompanied by a heavy quark which then decays to three jets (second diagram in Fig.~\ref{fig:axidecays}). This 4 jet final state would allow reconstruction of the total invariant mass as well as the heavy quark invariant mass at the Tevatron. At the LHC one would look for an 8 jet final state from axigluon pair production.

\item Decay to a scalar-pseudoscalar pair which each decay into 2 gluon jets (third diagram in Fig.~\ref{fig:axidecays}). This final state would allow reconstruction of both resonances as well as the total axigluon resonance at the Tevatron. At the LHC one would look for an 8 jet final state from axigluon pair production.

\item Decay to a pair of heavy fermions which decay into 3 jets each (fourth diagram in Fig.~\ref{fig:axidecays}). This final state resembles the decay products of hadronic top pairs. Similar events are also expected from R-parity violating gluino decays and a dedicated search for this final state was performed by CDF and CMS~\cite{Aaltonen:2011sg,Khachatryan:multijet}. In order to suppress the large QCD background both analyses applied very stringent cuts  which would eliminate all events in which the six jets come from axigluon decay. However, direct QCD pair production of the heavy quarks and subsequent decay to six jets would result in events which the search is sensitive to. This scenario is therefore strongly constrained by the two searches. The CDF search rules out heavy quark masses below about 140 GeV whereas the CMS search rules out masses from 170 GeV to about 300 GeV. These bounds apply to color octet fermions. Color triplets have smaller QCD cross sections, but most of our UV completions require multiple such fermions to obtain a sufficiently large axigluon width.

\end{enumerate}

\begin{acknowledgements}

We thank Nima Arkani-Hamed, Hsin-Chia Cheng, Clifford Cheung, Andy Cohen, Rouven Essig, Liam Fitzpatrick, David Lil' Kaplan, Jared Kaplan, Can Kilic, Zoltan Ligeti, Ann Nelson, Michael Peskin, Aaron Pierce, Tom Rizzo and Veronica Sanz for helpful conversations.
This research was supported in part by the U.S.\ Department of Energy under contract DE-FG02-01ER-40676 and by the National Science Foundation under Grant No. NSF PHY05-51164.

\end{acknowledgements}


\begin{thebibliography}{99}

\bibitem{CDFdilepton}
CDF Collaboration, CDF note 10436,
\url{http://www-cdf.fnal.gov/physics/new/top/2011/DilAfb/}.

\bibitem{Aaltonen:2008hc}
  T.~Aaltonen {\it et al.}  [CDF Collaboration],
  %``Forward-Backward Asymmetry in Top Quark Production in  $p\bar{p}$
  %Collisions at $sqrt{s}=1.96$ TeV,''
  Phys.\ Rev.\ Lett.\  {\bf 101}, 202001 (2008)
  [arXiv:0806.2472 [hep-ex]];
  %%CITATION = PRLTA,101,202001;%%
and CDF note 9724,
\url{http://www-cdf.fnal.gov/physics/new/top/2009/tprop/Afb/cdfnote_9724_public_v01.pdf}.

\bibitem{Abazov:2007qb}
  V.~M.~Abazov {\it et al.}  [D0 Collaboration],
  %``First measurement of the forward-backward charge asymmetry in top quark
  %pair production,''
  Phys.\ Rev.\ Lett.\  {\bf 100}, 142002 (2008)
  [arXiv:0712.0851 [hep-ex]].
  %%CITATION = PRLTA,100,142002;%%

\bibitem{Aaltonen:2011kc}
  T.~Aaltonen {\it et al.}  [CDF Collaboration],
  %``Evidence for a Mass Dependent Forward-Backward Asymmetry in Top Quark Pair
  %Production,''
  arXiv:1101.0034 [hep-ex].
  %%CITATION = ARXIV:1101.0034;%%

\bibitem{D0afb}
D\O\ Collaboration, D\O\ Note 6062-CONF,
\url{http://www-d0.fnal.gov/Run2Physics/WWW/results/prelim/TOP/T90/T90.pdf}.

\bibitem{neubert}
V.~Ahrens, A.~Ferroglia, M.~Neubert, B.~D.~Pecjak and L.~L.~Yang
%``The top-pair forward-backward asymmetry beyond NLO,''
 arXiv:1106.6051 [hep-ph].


\bibitem{Antunano:2007da}
  O.~Antunano, J.~H.~Kuhn and G.~Rodrigo,
  %``Top quarks, axigluons and charge asymmetries at hadron colliders,''
  Phys.\ Rev.\  D {\bf 77}, 014003 (2008)
  [arXiv:0709.1652 [hep-ph]].
  %%CITATION = PHRVA,D77,014003;%%

\bibitem{Bowen:2005ap}
  M.~T.~Bowen, S.~D.~Ellis and D.~Rainwater,
  %``Standard model top quark asymmetry at the Fermilab Tevatron,''
  Phys.\ Rev.\  D {\bf 73}, 014008 (2006)
  [arXiv:hep-ph/0509267].
  %%CITATION = PHRVA,D73,014008;%%

\bibitem{Kuhn:1998kw}
  J.~H.~Kuhn and G.~Rodrigo,
  %``Charge asymmetry of heavy quarks at hadron colliders,''
  Phys.\ Rev.\  D {\bf 59}, 054017 (1999)
  [arXiv:hep-ph/9807420].
  %%CITATION = PHRVA,D59,054017;%%

\bibitem{Almeida:2008ug}
  L.~G.~Almeida, G.~F.~Sterman and W.~Vogelsang,
  %``Threshold Resummation for the Top Quark Charge Asymmetry,''
  Phys.\ Rev.\  D {\bf 78}, 014008 (2008)
  [arXiv:0805.1885 [hep-ph]].
  %%CITATION = PHRVA,D78,014008;%%

%\cite{Frampton:2009rk}
\bibitem{Frampton:2009rk}
  P.~H.~Frampton, J.~Shu, K.~Wang,
  %``Axigluon as Possible Explanation for p anti-p ---> t anti-t Forward-Backward Asymmetry,''
  Phys.\ Lett.\  {\bf B683}, 294-297 (2010).
  [arXiv:0911.2955 [hep-ph]].


%\cite{Chivukula:2010fk}
\bibitem{Chivukula:2010fk}
  R.~S.~Chivukula, E.~H.~Simmons, C.~-P.~Yuan,
  %``Axigluons cannot explain the observed top quark forward-backward asymmetry,''
  Phys.\ Rev.\  {\bf D82}, 094009 (2010).
  [arXiv:1007.0260 [hep-ph]].

%\cite{Djouadi:2009nb}
\bibitem{Djouadi:2009nb}
  A.~Djouadi, G.~Moreau, F.~Richard, R.~K.~Singh,
  %``The Forward-backward asymmetry of top quark production at the Tevatron in warped extra dimensional models,''
  PHRVA,D82,071702.\ 2010 {\bf D82}, 071702 (2010).
  [arXiv:0906.0604 [hep-ph]].

%\cite{Djouadi:2011aj}
\bibitem{Djouadi:2011aj}
  A.~Djouadi, G.~Moreau, F.~Richard,
  %``Forward-backward asymmetries of the bottom and top quarks in warped extra-dimensional models: LHC predictions from the LEP and Tevatron anomalies,''
  Phys.\ Lett.\  {\bf B701}, 458-464 (2011).
  [arXiv:1105.3158 [hep-ph]].

%\cite{Bauer:2010iq}
\bibitem{Bauer:2010iq}
  M.~Bauer, F.~Goertz, U.~Haisch, T.~Pfoh, S.~Westhoff,
  %``Top-Quark Forward-Backward Asymmetry in Randall-Sundrum Models Beyond the Leading Order,''
  JHEP {\bf 1011}, 039 (2010).
  [arXiv:1008.0742 [hep-ph]].


%\cite{Alvarez:2010js}
\bibitem{Alvarez:2010js}
  E.~Alvarez, L.~Da Rold, A.~Szynkman,
  %``A composite Higgs model analysis of forward-backward asymmetries in the production of tops at Tevatron and bottoms at LEP and SLC,''
  JHEP {\bf 1105}, 070 (2011).
  [arXiv:1011.6557 [hep-ph]].


%\cite{Chen:2010hm}
\bibitem{Chen:2010hm}
  C.~-H.~Chen, G.~Cvetic, C.~S.~Kim,
  %``Forward-backward asymmetry of top quark in unparticle physics,''
  PHLTA,B694,393-397.\ 2011 {\bf B694}, 393-397 (2011).
  [arXiv:1009.4165 [hep-ph]].


%\cite{Bai:2011ed}
\bibitem{Bai:2011ed}
  Y.~Bai, J.~L.~Hewett, J.~Kaplan, T.~G.~Rizzo,
  %``LHC Predictions from a Tevatron Anomaly in the Top Quark Forward-Backward Asymmetry,''
  JHEP {\bf 1103}, 003 (2011).
  [arXiv:1101.5203 [hep-ph]].


%\cite{Zerwekh:2011wf}
\bibitem{Zerwekh:2011wf}
  A.~R.~Zerwekh,
  %``The Axigluon, a Four-Site Model and the Top Quark Forward-Backward Asymmetry at the Tevatron,''
 [arXiv:1103.0956 [hep-ph]].


%\cite{Barreto:2011au}
\bibitem{Barreto:2011au}
  E.~R.~Barreto, Y.~A.~Coutinho, J.~Sa Borges,
  %``Top quark forward-backward asymmetry from the $3-3-1$ model,''
  Phys.\ Rev.\  {\bf D83}, 054006 (2011).
  [arXiv:1103.1266 [hep-ph]].


%\cite{Jung:2009jz}
\bibitem{Jung:2009jz}
  S.~Jung, H.~Murayama, A.~Pierce, J.~D.~Wells,
  %``Top quark forward-backward asymmetry from new t-channel physics,''
  Phys.\ Rev.\  {\bf D81}, 015004 (2010).
  [arXiv:0907.4112 [hep-ph]].


%\cite{Cheung:2009ch}
\bibitem{Cheung:2009ch}
  K.~Cheung, W.~-Y.~Keung, T.~-C.~Yuan,
  %``Top Quark Forward-Backward Asymmetry,''
  Phys.\ Lett.\  {\bf B682}, 287-290 (2009).
  [arXiv:0908.2589 [hep-ph]].
  
  
  %\cite{Shu:2009xf}
\bibitem{Shu:2009xf}
  J.~Shu, T.~M.~P.~Tait, K.~Wang,
  %``Explorations of the Top Quark Forward-Backward Asymmetry at the Tevatron,''
  Phys.\ Rev.\  {\bf D81}, 034012 (2010).
  [arXiv:0911.3237 [hep-ph]].
  
  
  
  %\cite{Dorsner:2009mq}
\bibitem{Dorsner:2009mq}
  I.~Dorsner, S.~Fajfer, J.~F.~Kamenik, N.~Kosnik,
  %``Light colored scalars from grand unification and the forward-backward asymmetry in t t-bar production,''
  Phys.\ Rev.\  {\bf D81}, 055009 (2010).
  [arXiv:0912.0972 [hep-ph]].
  
  
  %\cite{Arhrib:2009hu}
\bibitem{Arhrib:2009hu}
  A.~Arhrib, R.~Benbrik, C.~-H.~Chen,
  %``Forward-backward asymmetry of top quark in diquark models,''
  Phys.\ Rev.\  {\bf D82}, 034034 (2010).
  [arXiv:0911.4875 [hep-ph]].
  
  %\cite{Barger:2010mw}
\bibitem{Barger:2010mw}
  V.~Barger, W.~-Y.~Keung, C.~-T.~Yu,
  %``Asymmetric Left-Right Model and the Top Pair Forward-Backward Asymmetry,''
  Phys.\ Rev.\  {\bf D81}, 113009 (2010).
  [arXiv:1002.1048 [hep-ph]].
  
  
  %\cite{Xiao:2010hm}
\bibitem{Xiao:2010hm}
  B.~Xiao, Y.~-k.~Wang, S.~-h.~Zhu,
  %``Forward-backward Asymmetry and Differential Cross Section of Top Quark in Flavor Violating Z' model at $\mathscr{O}(\alpha_s^2 \alpha_X)$,''
  Phys.\ Rev.\  {\bf D82}, 034026 (2010).
  [arXiv:1006.2510 [hep-ph]].
  
  
  %\cite{Gupta:2010wx}
\bibitem{Gupta:2010wx}
  S.~K.~Gupta,
  %``Same sign top-pairs in a non-universal Z' model at the LHC,''
   [arXiv:1011.4960 [hep-ph]].
  
 %\cite{Burdman:2010gr}
\bibitem{Burdman:2010gr}
  G.~Burdman, L.~de Lima and R.~D.~Matheus,
  %``New Strongly Coupled Sector at the Tevatron and the LHC,''
  Phys.\ Rev.\  D {\bf 83}, 035012 (2011)
  [arXiv:1011.6380 [hep-ph]].
  %%CITATION = PHRVA,D83,035012;%%
 
  %\cite{Cheung:2011qa}
\bibitem{Cheung:2011qa}
  K.~Cheung, T.~-C.~Yuan,
  %``Top Quark Forward-Backward Asymmetry in the Large Invariant Mass Region,''
  Phys.\ Rev.\  {\bf D83}, 074006 (2011).
  [arXiv:1101.1445 [hep-ph]].
  
  
  %\cite{Cao:2011ew}
\bibitem{Cao:2011ew}
  J.~Cao, L.~Wang, L.~Wu, J.~M.~Yang,
  %``Top quark forward-backward asymmetry, FCNC decays and like-sign pair production as a joint probe of new physics,''
    [arXiv:1101.4456 [hep-ph]].
  
  
  %\cite{Berger:2011ua}
\bibitem{Berger:2011ua}
  E.~L.~Berger, Q.~-H.~Cao, C.~-R.~Chen, C.~S.~Li, H.~Zhang,
  %``Top Quark Forward-Backward Asymmetry and Same-Sign Top Quark Pairs,''
  Phys.\ Rev.\ Lett.\  {\bf 106}, 201801 (2011).
  [arXiv:1101.5625 [hep-ph]].
  
  
  %\cite{Barger:2011ih}
\bibitem{Barger:2011ih}
  V.~Barger, W.~-Y.~Keung, C.~-T.~Yu,
  %``Tevatron Asymmetry of Tops in a W',Z' Model,''
  Phys.\ Lett.\  {\bf B698}, 243-250 (2011).
  [arXiv:1102.0279 [hep-ph]].
  
  
  %\cite{Grinstein:2011yv}
\bibitem{Grinstein:2011yv}
  B.~Grinstein, A.~L.~Kagan, M.~Trott, J.~Zupan,
  %``Forward-backward asymmetry in $t \bar{t}$t \bar{t} production from flavour symmetries,''
  Phys.\ Rev.\ Lett.\  {\bf 107}, 012002 (2011).
  [arXiv:1102.3374 [hep-ph]].
  
  
  %\cite{Patel:2011eh}
\bibitem{Patel:2011eh}
  K.~M.~Patel, P.~Sharma,
  %``Forward-backward asymmetry in top quark production from light colored scalars in SO(10) model,''
  JHEP {\bf 1104}, 085 (2011).
  [arXiv:1102.4736 [hep-ph]].
  
  
  %\cite{Jung:2009pi}
\bibitem{Jung:2009pi}
  D.~-W.~Jung, P.~Ko, J.~S.~Lee, S.~-H.~Nam,
  %``Model independent analysis of the forward-backward asymmetry of top quark production at the Tevatron,''
  Phys.\ Lett.\  {\bf B691}, 238-242 (2010).
  [arXiv:0912.1105 [hep-ph]].
  
  
  %\cite{Cao:2009uz}
\bibitem{Cao:2009uz}
  J.~Cao, Z.~Heng, L.~Wu, J.~M.~Yang,
  %``Top quark forward-backward asymmetry at the Tevatron: A Comparative study in different new physics models,''
  Phys.\ Rev.\  {\bf D81}, 014016 (2010).
  [arXiv:0912.1447 [hep-ph]].
  
  
  %\cite{Cao:2010zb}
\bibitem{Cao:2010zb}
  Q.~-H.~Cao, D.~McKeen, J.~L.~Rosner, G.~Shaughnessy, C.~E.~M.~Wagner,
  %``Forward-Backward Asymmetry of Top Quark Pair Production,''
  Phys.\ Rev.\  {\bf D81}, 114004 (2010).
  [arXiv:1003.3461 [hep-ph]].
  
  
  
  %\cite{Jung:2010yn}
\bibitem{Jung:2010yn}
  D.~-W.~Jung, P.~Ko, J.~S.~Lee,
  %``Longitudinal top polarization as a probe of a possible origin of forward-backward asymmetry of the top quark at the Tevatron,''
  PHLTA,B701,248-254.\ 2011 {\bf B701}, 248-254 (2011).
  [arXiv:1011.5976 [hep-ph]].
  
  
  %\cite{Choudhury:2010cd}
\bibitem{Choudhury:2010cd}
  D.~Choudhury, R.~M.~Godbole, S.~D.~Rindani, P.~Saha,
  %``Top polarization, forward-backward asymmetry and new physics,''
   [arXiv:1012.4750 [hep-ph]].
  
  
  %\cite{Jung:2010ri}
\bibitem{Jung:2010ri}
  D.~-W.~Jung, P.~Ko, J.~S.~Lee, S.~-H.~Nam,
  %``Model-independent analysis of forward-backward asymmetry of top quark production at the Tevatron,''
   [arXiv:1012.0102 [hep-ph]].
  
  
  %\cite{Blum:2011up}
\bibitem{Blum:2011up}
  K.~Blum, C.~Delaunay, O.~Gedalia, Y.~Hochberg, S.~J.~Lee, Y.~Nir, G.~Perez, Y.~Soreq,
  %``Implications of the CDF $t\bar{t}$ Forward-Backward Asymmetry for Boosted Top Physics,''
   [arXiv:1102.3133 [hep-ph]].
  
  
  
  %\cite{Craig:2011an}
\bibitem{Craig:2011an}
  N.~Craig, C.~Kilic, M.~J.~Strassler,
  %``LHC Charge Asymmetry as Constraint on Models for the Tevatron Top Anomaly,''
   [arXiv:1103.2127 [hep-ph]].
  
  
  %\cite{Delaunay:2011gv}
\bibitem{Delaunay:2011gv}
  C.~Delaunay, O.~Gedalia, Y.~Hochberg, G.~Perez, Y.~Soreq,
  %``Implications of the CDF $t \bar{t}$ Forward-Backward Asymmetry for Hard Top Physics,''
  [arXiv:1103.2297 [hep-ph]].
  
  
  %\cite{Foot:2011xu}
\bibitem{Foot:2011xu}
  R.~Foot,
  %``Top quark forward-backward asymmetry from SU($N_c$) color,''
  Phys.\ Rev.\  {\bf D83}, 114013 (2011).
  [arXiv:1103.1940 [hep-ph]].
  
  
  %\cite{Ligeti:2011vt}
\bibitem{Ligeti:2011vt}
  Z.~Ligeti, G.~M.~Tavares, M.~Schmaltz, 
  %``Explaining the t tbar forward-backward asymmetry without dijet or flavor anomalies,''
  JHEP {\bf 1106}, 109 (2011).
  [arXiv:1103.2757 [hep-ph]].
  
  
  %\cite{AguilarSaavedra:2011vw}
\bibitem{AguilarSaavedra:2011vw}
  J.~A.~Aguilar-Saavedra, M.~Perez-Victoria,
  %``Probing the Tevatron t tbar asymmetry at LHC,''
  JHEP {\bf 1105}, 034 (2011).
  [arXiv:1103.2765 [hep-ph]].
  
  
  %\cite{Jung:2011zv}
\bibitem{Jung:2011zv}
  S.~Jung, A.~Pierce, J.~D.~Wells,
  %``Top quark asymmetry from a non-Abelian horizontal symmetry,''
   [arXiv:1103.4835 [hep-ph]].
  
  
  %\cite{Hewett:2011wz}
\bibitem{Hewett:2011wz}
  J.~L.~Hewett, J.~Shelton, M.~Spannowsky, T.~M.~P.~Tait, M.~Takeuchi,
  %``$A^t_{FB}$ Meets LHC,''
   [arXiv:1103.4618 [hep-ph]].
  
  
  %\cite{Nelson:2011us}
\bibitem{Nelson:2011us}
  A.~E.~Nelson, T.~Okui, T.~S.~Roy,
  %``A unified, flavor symmetric explanation for the t-tbar asymmetry and Wjj excess at CDF,''
   [arXiv:1104.2030 [hep-ph]].
  
  
  %\cite{Krohn:2011tw}
\bibitem{Krohn:2011tw}
  D.~Krohn, T.~Liu, J.~Shelton, L.~-T.~Wang,
  %``A Polarized View of the Top Asymmetry,''
   [arXiv:1105.3743 [hep-ph]].
  
  
  %\cite{AguilarSaavedra:2011hz}
\bibitem{AguilarSaavedra:2011hz}
  J.~A.~Aguilar-Saavedra, M.~Perez-Victoria,
  %``Asymmetries in t $\bar{t}$ production: LHC versus Tevatron,''
   [arXiv:1105.4606 [hep-ph]].
  
%\cite{Haisch:2011up}
\bibitem{Haisch:2011up}
  U.~Haisch, S.~Westhoff,
  %``Massive Color-Octet Bosons: Bounds on Effects in Top-Quark Pair Production,''
   [arXiv:1106.0529 [hep-ph]].  
  
  %\cite{Cui:2011xy}
\bibitem{Cui:2011xy}
  Y.~Cui, Z.~Han, M.~D.~Schwartz,
  %``Top condensation as a motivated explanation of the top forward-backward asymmetry,''
  [arXiv:1106.3086 [hep-ph]].
  

  %\cite{Barcelo:2011vk}
\bibitem{Barcelo:2011vk}
  R.~Barcelo, A.~Carmona, M.~Masip, J.~Santiago,
  %``Stealth gluons at hadron colliders,''
   [arXiv:1106.4054 [hep-ph]].
  
  
  %\cite{Gabrielli:2011jf}
\bibitem{Gabrielli:2011jf}
  E.~Gabrielli, M.~Raidal,
  %``Effective axial-vector coupling of gluon as an explanation to the top quark asymmetry,''
   [arXiv:1106.4553 [hep-ph]].
  
  
  %\cite{Duraisamy:2011pt}
\bibitem{Duraisamy:2011pt}
  M.~Duraisamy, A.~Rashed, A.~Datta,
  %``The Top Forward Backward Asymmetry with general Z ' couplings,''
   [arXiv:1106.5982 [hep-ph]].
  

\bibitem{Shelton:2011hq}
  J.~Shelton and K.~M.~Zurek,
  %``A Theory for Maximal Flavor Violation,''
  arXiv:1101.5392 [hep-ph].
  %%CITATION = ARXIV:1101.5392;%%

\bibitem{Degrande:2010kt}
  C.~Degrande, J.~M.~Gerard, C.~Grojean, F.~Maltoni and G.~Servant,
  %``Non-resonant New Physics in Top Pair Production at Hadron Colliders,''
  arXiv:1010.6304 [hep-ph].
  %%CITATION = ARXIV:1010.6304;%%

%\cite{AguilarSaavedra:2011ug}
\bibitem{AguilarSaavedra:2011ug}
  J.~A.~Aguilar-Saavedra and M.~Perez-Victoria,
  %``Simple models for the top asymmetry: constraints and predictions,''
  arXiv:1107.0841 [hep-ph].
  %%CITATION = ARXIV:1107.0841;%%


\bibitem{Hall:1985wz}
  L.~J.~Hall and A.~E.~Nelson,
  %``HEAVY GLUONS AND MONOJETS,''
  Phys.\ Lett.\  B {\bf 153}, 430 (1985).
  %%CITATION = PHLTA,B153,430;%%

\bibitem{Frampton:1987dn}
  P.~H.~Frampton and S.~L.~Glashow,
  %``Chiral Color: An Alternative to the Standard Model,''
  Phys.\ Lett.\  B {\bf 190}, 157 (1987).
  %%CITATION = PHLTA,B190,157;%%

\bibitem{Bagger:1987fz}
  J.~Bagger, C.~Schmidt and S.~King,
  %``AXIGLUON PRODUCTION IN HADRONIC COLLISIONS,''
  Phys.\ Rev.\  D {\bf 37}, 1188 (1988).
  %%CITATION = PHRVA,D37,1188;%%

\bibitem{CDFsigma}
CDF Collaboration, CDF note 9913,
\url{http://www-cdf.fnal.gov/physics/new/top/confNotes/cdf9913_ttbarxs4invfb.ps}.

\bibitem{Aaltonen:2009iz}
  T.~Aaltonen {\it et al.}  [CDF Collaboration],
  %``First Measurement of the $t\bar{t}$ Differential Cross Section
  %${d\sigma/d}M_{t\overline{t}}$ in $p\bar{p}$ Collisions at $\sqrt{s}=1.96$
  %TeV,''
  Phys.\ Rev.\ Lett.\  {\bf 102}, 222003 (2009)
  [arXiv:0903.2850 [hep-ex]].
  %%CITATION = PRLTA,102,222003;%%


\bibitem{Moch:2008qy}
  S.~Moch and P.~Uwer,
  %``Theoretical status and prospects for top-quark pair production at hadron
  %colliders,''
  Phys.\ Rev.\  D {\bf 78}, 034003 (2008)
  [arXiv:0804.1476 [hep-ph]].
  %%CITATION = PHRVA,D78,034003;%%

\bibitem{Cacciari:2008zb}
  M.~Cacciari, S.~Frixione, M.~L.~Mangano, P.~Nason and G.~Ridolfi,
  %``Updated predictions for the total production cross sections of top and of
  %heavier quark pairs at the Tevatron and at the LHC,''
  JHEP {\bf 0809}, 127 (2008)
  [arXiv:0804.2800 [hep-ph]].
  %%CITATION = JHEPA,0809,127;%%

\bibitem{Kidonakis:2008mu}
  N.~Kidonakis and R.~Vogt,
  %``The Theoretical top quark cross section at the Tevatron and the LHC,''
  Phys.\ Rev.\  D {\bf 78}, 074005 (2008)
  [arXiv:0805.3844 [hep-ph]].
  %%CITATION = PHRVA,D78,074005;%%

\bibitem{Ahrens:2010zv}
  V.~Ahrens, A.~Ferroglia, M.~Neubert, B.~D.~Pecjak and L.~L.~Yang,
  %``Renormalization-Group Improved Predictions for Top-Quark Pair Production at
  %Hadron Colliders,''
  JHEP {\bf 1009}, 097 (2010)
  [arXiv:1003.5827 [hep-ph]];
  %%CITATION = JHEPA,1009,097;%%
%\bibitem{Ahrens:2011mw}
%  V.~Ahrens, A.~Ferroglia, M.~Neubert, B.~D.~Pecjak and L.~L.~Yang,
  %``RG-improved single-particle inclusive cross sections and forward-backward
  %asymmetry in $t\bar t$ production at hadron colliders,''
  arXiv:1103.0550 [hep-ph].
  %%CITATION = ARXIV:1103.0550;%%


%\cite{Chatrchyan:2011ew}
\bibitem{Chatrchyan:2011ew}
  S.~Chatrchyan {\it et al.} [ CMS Collaboration ],
  %``Measurement of the Top-antitop Production Cross Section in pp Collisions at sqrt(s)=7 TeV using the Kinematic Properties of Events with Leptons and Jets,''
   [arXiv:1106.0902 [hep-ex]].

\bibitem{atlasttbar}
  G.~Aad {\it et al.}  [ATLAS Collaboration],
%``A Search for ttbar Resonances in the Lepton Plus Jets Channel using 200/pb of pp Collisions at √s = 7 TeV
ATLAS-CONF-2011-087,
http://cdsweb.cern.ch/record/1356196.

\bibitem{Aaltonen:2008dn}
  T.~Aaltonen {\it et al.}  [CDF Collaboration],
  %``Search for new particles decaying into dijets in proton-antiproton
  %collisions at sqrt(s) = 1.96 TeV,''
  Phys.\ Rev.\  D {\bf 79}, 112002 (2009)
  [arXiv:0812.4036 [hep-ex]].
  %%CITATION = PHRVA,D79,112002;%%

\bibitem{CDFchidist}
CDF Collaboration, CDF Note 9609, 
\url{http://www-cdf.fnal.gov/physics/new/qcd/dijetchi_08/}.

\bibitem{Abazov:2009mh}
  V.~M.~Abazov {\it et al.}  [D0 Collaboration],
  %``Measurement of dijet angular distributions at sqrt{s}=1.96TeV and searches
  %for quark compositeness and extra spatial dimensions,''
  Phys.\ Rev.\ Lett.\  {\bf 103}, 191803 (2009)
  [arXiv:0906.4819 [hep-ex]].
  %%CITATION = PRLTA,103,191803;%%

\bibitem{Khachatryan:2011as}
  V.~Khachatryan {\it et al.}  [CMS Collaboration],
  %``Measurement of Dijet Angular Distributions and Search for Quark
  %Compositeness in pp Collisions at 7 TeV,''
  arXiv:1102.2020 [hep-ex].
  %%CITATION = ARXIV:1102.2020;%%

\bibitem{Khachatryan:2010jd}
  V.~Khachatryan {\it et al.}  [CMS Collaboration],
  %``Search for Dijet Resonances in 7 TeV pp Collisions at CMS,''
  Phys.\ Rev.\ Lett.\  {\bf 105}, 211801 (2010)
  [arXiv:1010.0203 [hep-ex]].
  %%CITATION = PRLTA,105,211801;%%

\bibitem{Collaboration:2010eza}
  G.~Aad {\it et al.}  [ATLAS Collaboration],
  %``Search for Quark Contact Interactions in Dijet Angular Distributions in pp
  %Collisions at sqrt(s) = 7 TeV Measured with the ATLAS Detector,''
  Phys.\ Lett.\  B {\bf 694}, 327 (2011)
  [arXiv:1009.5069 [hep-ex]].
  %%CITATION = PHLTA,B694,327;%%
 
%\cite{Fox:2011qd}
\bibitem{Fox:2011qd}
  P.~J.~Fox, J.~Liu, D.~Tucker-Smith, N.~Weiner,
  %``An Effective Z',''
  [arXiv:1104.4127 [hep-ph]].

%\cite{Kilic:2008ub}
\bibitem{Kilic:2008ub}
  C.~Kilic, S.~Schumann, M.~Son,
  %``Searching for Multijet Resonances at the LHC,''
  JHEP {\bf 0904}, 128 (2009).
  [arXiv:0810.5542 [hep-ph]].
  
  %\cite{Dicus:2010bm}
\bibitem{Dicus:2010bm}
  D.~A.~Dicus, C.~Kao, S.~Nandi, J.~Sayre,
  %``Discovering Colorons at the Early Stage LHC,''
  Phys.\ Rev.\  {\bf D83}, 091702 (2011).
  [arXiv:1012.5694 [hep-ph]].

  %\cite{4bees}
\bibitem{4bees}
CDF Collaboration, CDF Note 10105, 
%``Search for Higgs Bosons Produced in Association with b-Quarks''
\url{http://www-cdf.fnal.gov/physics/new/hdg//Results_files/results/3b_susyhiggs_jun10/cdf10105_higgs3b_public.pdf}.

%\cite{Abazov:2010ci}
\bibitem{Abazov:2010ci}
  V.~M.~Abazov {\it et al.}  [D0 Collaboration],
  %``Search for neutral Higgs bosons in the multi-$b$-jet topology in
  %5.2fb$^{-1}$ of $p\bar{p}$ collisions at $\sqrt{s} = 1.96$ TeV,''
  Phys.\ Lett.\  B {\bf 698}, 97 (2011)
  [arXiv:1011.1931 [hep-ex]].
  %%CITATION = PHLTA,B698,97;%%


%\cite{Essig:2008zz}
\bibitem{Essig:2008zz}
  R.~Essig,
  %``Physics beyond the standard model: Supersymmetry, dark matter, and LHC phenomenology,''
Ph.D. thesis, AAT-3349692, PROQUEST-1692394711. Oct 2008.
  

%\cite{Aaltonen:2011sg}
\bibitem{Aaltonen:2011sg}
  T.~Aaltonen {\it et al.} [ CDF Collaboration ],
  %``First Search for Multijet Resonances in $\sqrt{s} = 1.96$ TeV $ p\bar{p}$ Collisions,''
   [arXiv:1105.2815 [hep-ex]].

\bibitem{Khachatryan:multijet}
  V.~Khachatryan {\it et al.}  [CMS Collaboration],
%``Search for Multijet Resonances in pp Collisions at sqrt s = 7 TeV''
\url{http://cdsweb.cern.ch/record/1351471?ln=en}.

%\cite{Alves:2010za}
\bibitem{Alves:2010za}
  D.~S.~M.~Alves, E.~Izaguirre, J.~G.~Wacker,
  %``It's On: Early Interpretations of ATLAS Results in Jets and Missing Energy Searches,''
    [arXiv:1008.0407 [hep-ph]].

\end{thebibliography}
\end{document}